\documentclass[11pt]{amsart}
\newcommand{\transpo}{{}^t\!}

\newtheorem{Theorem}{Theorem}[section]

\newtheorem{Lemma}[Theorem]{Lemma}

\newtheorem{Assumption 2}[Theorem]{Assumption 2}

\numberwithin{equation}{section}

\usepackage{amsmath}
\usepackage{amssymb}
\usepackage{lmodern}
\usepackage{stmaryrd}
\usepackage{ulem}
\usepackage{yfonts}
\usepackage{amsbsy}
\usepackage{pstricks}
\usepackage{graphicx}
\usepackage{pst-xkey}
\usepackage{pst-all,pst-gr3d}
\usepackage[dvips,arrow,matrix]{xy}
\usepackage{gensymb}
\usepackage{tikz}

\def\k#1{\kern#1em}
\def\Ib#1{{I\kern-.25em#1}}
\def\Ibb#1{{I\kern-.23em#1}}

\def\CC{{\mathbb C}}

\def\NN{{\mathbb N}}

\def\RR{{\mathbb{R}}}

\def\vci{\vrule  width.02em height1.47ex depth-.0ex}
\def\11{{\rm\k{.2}\vci\k{-.37}1}}

\def\fin{{\begin{flushright}
\it{Q.E.D.}
\end{flushright}}}

\begin{document}

\address{Universit\'e de Bordeaux, Institut de Math\'ematiques, UMR CNRS 5251, F-33405 Talence Cedex}

\email{bachelot@math.u-bordeaux1.fr}

\title{The Klein-Gordon equation in the Anti-de Sitter
  Cosmology}

\author{Alain BACHELOT}

\begin{abstract}
This paper deals with the Klein-Gordon equation on the Poincar\'e chart of the 5-dimensional Anti-de Sitter universe. 
When the mass $\mu$ is larger than $-\frac{1}{4}$, the Cauchy problem is well posed despite the loss of global hyperbolicity due to the time-like horizon. We express the finite energy solutions in the form of a continuous Kaluza-Klein tower and we deduce a uniform decay as $\mid t\mid^{-\frac{3}{2}}$.
We investigate the case $\mu=\frac{\nu^2-1}{2}$, $\nu\in\NN^*$, which encompasses the gravitational fluctuations, $\nu=4$, and the electromagnetic waves, $\nu=2$. The propagation of the wave front set shows that the horizon acts like a perfect mirror. We establish that the smooth solutions decay as  $\mid t\mid^{-2-\sqrt{\mu+\frac{1}{4}}}$, and we get global $L^p$ estimates of Strichartz type. When $\nu$ is even, there appears a lacuna and the equipartition of the energy occurs at finite time for the compactly supported initial data, although the Huygens principle fails. We adress the cosmological model of the negative tension Minkowski brane, on which a Robin boundary condition is imposed. We prove the hyperbolic mixed problem is well-posed and the normalizable solutions can be expanded into a discrete Kaluza-Klein tower. We establish some $L^2-L^{\infty}$ estimates in suitable weighted Sobolev spaces.
\end{abstract}

\maketitle


\pagestyle{myheadings}
\markboth{\centerline{\sc Alain Bachelot}}{\centerline{\sc The Klein-Gordon
    equation in Anti-de Sitter Cosmology}}

\section{Introduction}
The Anti-de-Sitter space-time $AdS^{n+1}$,  is the unique $n+1$-dimensional maximally symmetric solution without singularity, of the Einstein equations in the vacuum, with a negative cosmological constant (see the Appendix for a brief presentation). The recent surge of interest about $AdS^5$ came from the hope to construct a quantum gravity from a string theory. This dream has aroused a huge literature by the physicists, but there are few mathematical works dealing with the PDE of the fields theory in this geometrical framework. The problems arising in the context of $AdS^{n+1}$ are somewhat unusual because of the crucial property of this lorentzian manifold : the loss of global hyperbolicity due to the existence of a time-like horizon. There are also closed time-like curves, but this unpleasant property disappears when we consider the universal covering $CAdS^{n+1}$,  hence the main issue consists in understanding the role of the horizon in the existence, the uniqueness, and the qualitative properties of the solutions of the wave equations. 
We have investigated the Dirac system in $CAdS^4$ in \cite{DADS} and the wave equations associated to each spherical harmonics of the scalar/electromagnetic/gravitational fluctuations on the whole $CAdS^{n+1}$ has been studied by A. Ishibashi and  R. M. Wald \cite{ishi2}. The scalar waves on $CAdS^4$ were discussed in \cite{avis} and \cite{breit1}.\\

In this paper we consider the Klein-Gordon equation
in the Poincar\'e patch of $AdS^5$. It is a strictly included subdomain $\mathcal{P}$ of $AdS^5$ that plays a fundamental role in brane cosmology (see e.g. \cite{man}). This manifold is defined by
\begin{equation}
\mathcal{P}:=\RR_t\times\RR_{\mathbf x}^3\times]0,\infty[_z,\;\;
g_{\mu\nu}d^{\mu}dx^{\nu}=\left(\frac{1}{kz}\right)^2\left(dt^2-d{\mathbf x}^2-dz^2\right).
 \label{poincare
}
\end{equation}
The boundary of this universe, that is located at $z=0$, is time-like and we can see that many null geodesics hit this horizon, so $\mathcal{P}$ is not globally hyperbolic. The main aim of this paper consists in the understanding of the role of this horizon related to the propagation of the fields and their asymptotic behaviours. An important motivation for our work is the study of the gravitational waves in this geometry that are some fluctuations $u_{ij}$ of the metric $AdS^5$
$$
 ds^2=\left(\frac{1}{kz}\right)^2\left(dt^2-(\delta_{ij}+u_{ij})dx^idx^j-dz^2\right).
$$
They are solutions of the linearized Einstein equations, which are simply reduced to the D'Alembertian in $AdS^5$
$$
\square_g u=0,\;\;\;
\square_g:=\frac{1}{\sqrt{\mid g\mid}}\frac{\partial}{\partial
  x^{\mu}}\left(\sqrt{\mid g\mid}g^{\mu\nu}\frac{\partial}{\partial
  x^{\nu}}\right).
$$
More generally, we consider the Klein-Gordon equation
\begin{equation}
\square_gu+\lambda k^2u=0,
  \label{kgo}
\end{equation}
with $\lambda\in\RR$. In fact this equation is a master equation that appears for other fields : for the scalar fields, $\lambda$ is the mass, for the vector electromagnetic fields $\lambda=-3$. If we put $\Phi=:z^{-\frac{3}{2}}u$ and $\mu:=\frac{15}{4}+\lambda$, the equation (\ref{kgo}) on $\mathcal{P}$ takes the very simple form of the free wave equation on the 1+4-dimensional half Minkowski space-time $\RR_t\times\RR^3_{\mathrm x}\times]0,\infty[_z$, pertubed by a singular cartesian potential $\frac{\mu}{z^2}$:
\begin{equation}
\left(\partial_t^2-\Delta_{\mathbf{x}}-\partial_z^2+\frac{\mu}{z^2}\right)\Phi=0.
  \label{eq}
\end{equation}

Since $\mathcal{P}$ is not globally hyperbolic, the well-posedness of the Cauchy problem is doubtful and the question arises of the necessity to impose some boundary condition on the time-like horizon. Nevertheless, we can establish that no time-like geodesic hits this boundary, therefore we may hope that the Cauchy problem is well-posed for the wave equations when the mass of the field is large enough. In the following part, we show that this is indeed the case when $\mu>-\frac{1}{4}$ (i.e. $\lambda>-4$), and there exists a unique solution of the initial value problem when the natural energy associated with (\ref{eq}) is finite. In fact this constraint is equivalent to the Dirichlet condition on the horizon. We also prove that the solutions are a superposition of a continuum Klein-Gordon fields (the so called Kaluza-Klein tower). We deduce that the smooth solutions decay uniformly as $\mid t\mid^{-\frac{3}{2}}$ and behave near the horizon as $z^{-\frac{1}{2}-\sqrt{\mu+\frac{1}{4}}}$.

In part 3, we obtain more precise properties when the mass $\mu$ has the form $\mu=\frac{\nu^2-1}{4}$, $\nu\in\NN^*$. In this case (\ref{eq}) is closely linked with the wave equation in a higher dimension Minkowski space. This spectrum of mass includes the gravitational waves ($\mu=\frac{15}{4}$), and the electromagnetic waves ($\mu=\frac{3}{4}$). The Huygens principle fails for (\ref{eq}), but a lacuna appears, and the equipartition of the energy occurs at finite time for the compactly supported data, when $\nu$ is even. Moreover for all $\nu$, the horizon acts like a perfect miror : the singularities are reflected according to the Descartes law. We prove a strong decay of the  smooth solutions that behave as  $\mid t\mid^{-2-\sqrt{\mu+\frac{1}{4}}}z^{-\frac{1}{2}-\sqrt{\mu+\frac{1}{4}}}$ We establish also global $L^p$ space-time estimates of Strichartz type in weighted spaces for the finite energy solutions.

In the fourth part, we adress the cosmological problem of the Minkowski brane (see e.g. \cite{man}).  This brane is just the submanifold $\RR_t\times\RR^3_{\mathrm x}\times\{z=1\}$. In \cite{RS} we have developed a complete analysis of the gravitational fluctuations for the positive-tension Minkowski brane that is the boundary of $\RR_t\times\RR^3_{\mathrm x}\times]1,\infty[_z$. In this paper, we investigate the Klein-Gordon equation for the negative-tension Minkowski brane considered as the part $z=1$ of the boundary of the Anti-de Sitter bulk $\mathcal{B}:=\RR_t\times\RR^3_{\mathrm x}\times]0,1[_z$.  The dynamics of the field is given by the equation (\ref{eq}) in $\mathcal{B}$ and the Neumann condition on the brane, $\partial_zu=0$, that is equivalent to
\begin{equation}
\partial_z\Phi(t,\mathrm{x},1)+\frac{3}{2}\Phi(t,\mathrm{x},1)=0.
 \label{}
\end{equation}
We solve the mixed problem and expand the normalizable solutions in the form of a discrete Kaluza-Klein tower ; this result provides a rigorous  functional framework to the expansions of the physicists. We establish some $L^2-L^{\infty}$ estimates that express that the smooth fields decay as
$t^{-\frac{3}{2}}z^{-\frac{1}{2}-\sqrt{\mu+\frac{1}{4}}}$ again. Finally a short appendix presents the basics elements of the $AdS^5$ brane cosmology.

\section{Finite energy solutions on $AdS^5$}
In this part we investigate the finite energy  solutions of the Klein-Gordon equation (\ref{eq})
where $\mu$ is a real number satisfying
\begin{equation}
-\frac{1}{4}<\mu.
  \label{}
\end{equation}
We shall see that this constraint is natural to assure the positivity of the energy
that is the formally conserved quantity associated with the equation
(\ref{eq}) :
\begin{equation}
E(\Phi,t):=\int_{\RR^3}\int_0^{\infty}\mid\nabla_{t,{\mathbf x},z}\Phi(t,{\mathbf
  x},z)\mid^2+\frac{\mu}{z^2}\mid\Phi(t,{\mathbf x},z)\mid^2d{\mathbf x}dz
  \label{ener}
\end{equation}
The solutions $\Phi$ of (\ref{eq}) such that for all time $t$
\begin{equation}
0\leq E(\Phi,t)<\infty
  \label{enerf}
\end{equation}
are
called {\it finite energy solutions}. We prove that under this
constraint, {\it i.e.} in a suitable functional framework, there is a
unique dynamics associated to equation (\ref{eq}).

To ensure that  the energy (\ref{ener}) is finite, it is natural to
impose that for all fixed time $t$, $\Phi(t,.)$ belongs
to $H^1(\RR^3_{\mathbf x}\times]0,1[_z)$. Since $\Phi(t,.)\in
C^0([0,1]_z;H^{\frac{1}{2}}(\RR^3_{\mathbf x}))$, the constraint
$\frac{1}{z}\Phi(t,.)\in L^2(\RR^3_{\mathbf x}\times]0,1[_z)$ implies
that $\Phi(t,\mathbf{x}, 0)=0$. Hence the finiteness of the energy implicitely 
requires the Dirichlet condition on the time-like horizon $z=0$ of the
anti-de Sitter space-time. Now we construct the functional framework
associated to the energy.
Following J. Deny and J-L. Lions \cite{deny}, given an open set $\Omega\subset\RR^N_{\mathbf z}$ and a Hilbert space $X$, we introduce the Beppo Levi
space $BL^1_0(\Omega;X)$ (respectively $BL^1(\Omega;X)$)  as the completion of the space of the test
functions $C^{\infty}_0(\Omega;X)$ (respectively
$C^{\infty}_0(\overline{\Omega} ;X)$) for the norm
$\|\nabla.\|_{L^2(\Omega;X)}$ (and we omit $X$ when $X=\CC$). When
$N\geq 3$ these spaces are sets of distributions and we have
\begin{equation}
\int_{\Omega}\frac{\mid\phi(\mathbf{z})\mid_X^2}{\mid\mathbf{z}\mid^2}d\mathbf{z}\leq\frac{4}{(N-2)^2}\int_{\Omega}\mid\nabla_{\mathbf{z}}\phi(\mathbf{z})\mid_X^2d\mathbf{z}
  \label{}
\end{equation}
and also
\begin{equation}
BL^1_0(\Omega)\subset L^{\frac{2N}{N-2}}(\Omega,d\mathbf{z}).
  \label{}
\end{equation}
Thanks to the Hardy inequality
\begin{equation}
\forall\phi\in C^{\infty}_0(]0,\infty[;X),\;\;\forall A>0,\;\;\int_0^A\frac{1}{z^2}\left\vert\int_0^z
  \phi(\zeta)d\zeta\right\vert_X^2dz\leq 4\int_0^A\left\vert \phi(z)\right\vert_X^2dz
  \label{hardy}
\end{equation}
we see that for all $\phi\in BL^1_0(\RR^3_{\mathbf x}\times
]0,\infty[_z)$, we have :
\begin{equation}
\int_{\RR^3}\int_0^{\infty}\mid\partial_z\phi({\mathbf x},z)\mid^2
+\frac{\mu}{z^2}\mid\phi({\mathbf x},z)\mid^2d{\mathbf x}dz
\leq (1+4\mid \mu\mid)\int_{\RR^3}\int_0^{\infty}\mid\partial_z\phi({\mathbf x},z)\mid^2d{\mathbf x}dz
  \label{}
\end{equation}
and when $-\frac{1}{4}<\mu\leq 0$
\begin{equation}
 (1+4\mu)\int_{\RR^3}\int_0^{\infty}\mid\partial_z\phi({\mathbf
   x},z)\mid^2d{\mathbf x}dz\leq \int_{\RR^3}\int_0^{\infty}\mid\partial_z\phi({\mathbf x},z)\mid^2
+\frac{\mu}{z^2}\mid\phi({\mathbf x},z)\mid^2d{\mathbf x}dz
  \label{}
\end{equation}
hence (\ref{enerf}) is satisfied by  the solutions $\Phi$ of
(\ref{eq}) such that
\begin{equation}
\Phi\in C^0\left(\RR_t;BL^1_0\left(\RR^3_{\mathbf
    x}\times]0,\infty[_z\right)\right),\;\;
\partial_t\Phi\in C^0\left(\RR_t;L^2\left(\RR^3_{\mathbf
    x}\times]0,\infty[_z\right)\right).
  \label{reg}
\end{equation}

In the sequel, 
$BL^1_0\left(\RR^3_{\mathbf
    x}\times]0,\infty[_z\right)$ is endowed with the norm
\begin{equation}
\|\Phi\|_{BL^1_0}^2:=\int_{\RR^3}\int_0^{\infty}\mid\nabla_{{\mathbf x},z}\Phi({\mathbf
  x},z)\mid^2+\frac{\mu}{z^2}\mid\Phi({\mathbf x},z)\mid^2d{\mathbf
  x}dz.
  \label{norm}
\end{equation}
It will be useful to note that
\begin{equation}
\|\Phi\|_{BL^1_0}^2=\int_{\RR^3}\int_0^{\infty}\mid\nabla_{{\mathbf x}}\Phi({\mathbf
  x},z)\mid^2+\left\vert\partial_z\Phi({\mathbf
  x},z)+\frac{\alpha_{\pm}}{z}\Phi({\mathbf
  x},z)\right\vert^2d{\mathbf
  x}dz,
  \label{normz}
\end{equation}
with 
\begin{equation}
\alpha_{\pm}=-\frac{1}{2}\pm\sqrt{\mu+\frac{1}{4}}.
  \label{alfa}
\end{equation}
We are ready to state the result of existence of the finite energy solutions.
\begin{Theorem}
For any $\mu>-\frac{1}{4}$, 
given $\Phi_0\in BL^1_0\left(\RR^3_{\mathbf
    x}\times]0,\infty[_z\right)$ and $\Phi_1\in L^2\left(\RR^3_{\mathbf
    x}\times]0,\infty[_z\right)$, there exists a unique solution $\Phi$
of (\ref{eq}) satisfying (\ref{reg}) and the Cauchy condition
\begin{equation}
\Phi(0,{\mathbf x},z)=\Phi_0({\mathbf x},z),\;\;\partial_t\Phi(0,{\mathbf x},z))=\Phi_1({\mathbf x},z),\;\;({\mathbf x},z)\in\RR^3\times]0,\infty[.
  \label{ci}
\end{equation}
Moreover the energy (\ref{ener}) is conserved :
\begin{equation}
\forall t\in\RR,\;\;E(\Phi,t)=E(\Phi,0),
  \label{cons}
\end{equation}
and if $\Phi_0(\mathbf{x},z)=\Phi_1(\mathbf{x},z)=0$ when $\mid \mathbf{x}\mid\geq R$ or $\mid z\mid\geq R$, then $\Phi(t,\mathbf{x},z)=0$ when $\mid \mathbf{x}\mid\geq R+\mid t \mid$ or $\mid z\mid\geq R+\mid t\mid$.
  \label{theoexist}
\end{Theorem}

{\it Proof of Theorem\ref{theoexist}.}
We introduce the densely defined operator on $BL^1_0\left(\RR^3_{\mathbf
    x}\times]0,\infty[_z\right)\times L^2\left(\RR^3_{\mathbf
    x}\times]0,\infty[_z\right)$, given by 
$$
A_0:=\frac{1}{i}\left(
\begin{array}{cc}
0&1\\
\Delta_{\mathbf{x}}+\partial_z^2-\frac{\mu}{z^2}&0
\end{array}
\right),\;\;Dom(A_0)=C^{\infty}_0\left(\RR^3_{\mathbf
    x}\times]0,\infty[_z\right)\times C^{\infty}_0\left(\RR^3_{\mathbf
    x}\times]0,\infty[_z\right).
$$
$A_0$ is obviously symmetric. Now we prove that it is essentially self-adjoint.
We easily show that its adjoint $A_0^*$ is defined by
$$
A_0^*:=\frac{1}{i}\left(
\begin{array}{cc}
0&1\\
\Delta_{\mathbf{x}}+\partial_z^2-\frac{\mu}{z^2}&0
\end{array}
\right),\;\;Dom(A_0^*)=\left\{\Phi_0\in BL^1_0;\;-\Delta_{\mathbf{x},z}\Phi_0+\frac{\mu}{z^2}\Phi_0\in L^2\right\}\times H^1_0,
$$
where $H^1_0=BL^1_0\cap L^2$ is the usual Sobolev space. We consider
$(\Phi_0^{\pm},\Phi_1^{\pm})\in Ker(A_0^*\pm i)$. Then
$\Phi_1^{\pm}=\pm \Phi_0^{\pm}\in H^1_0$ and 
$$
\left[-\Delta_{\mathbf{x},z}+\frac{\mu}{z^2}+1\right]\Phi_0^{\pm}=0.
$$
Since $\phi\mapsto \Delta_{\mathbf{x},z}\phi$ and $\phi\mapsto
\frac{1}{z^2}\phi$ are bounded from
$H^1_0\left(\RR^3_{\mathbf x}\times]0,\infty[_z\right)$ to its dual
space $H^{-1}\left(\RR^3_{\mathbf x}\times]0,\infty[_z\right)$, we deduce that
$$
\|\Phi_0^{\pm}\|_{BL^1_0}^2+\|\Phi_0^{\pm}\|_{L^2}^2=\left<\left[-\Delta_{\mathbf{x},z}+\frac{\mu}{z^2}+1\right]\Phi_0^{\pm};\Phi_0^{\pm}\right>_{H^{-1},H^1_0}=0,
$$
hence we conclude that $\Phi_j^{\pm}=0$ and $A_0$ is essentially
self-adjoint.
If $A$ is its unique self-adjoint extension, $\Phi(t,.)$ given by 
$(\Phi(t),\partial_t\Phi(t))=e^{itA}(\Phi_0,\Phi_1)$ satisfies
(\ref{eq}), (\ref{reg}), (\ref{ci}) and (\ref{cons}). The uniqueness is
established by a classical way. Given a solution $\Phi$ of (\ref{eq}),
(\ref{reg}), (\ref{ci}), we put for all $\varepsilon>0$,
$\Phi_{\varepsilon}(t)=\frac{1}{\varepsilon}\int_t^{t+\varepsilon}\Phi(\tau)d\tau$.
Then $\Phi_{\varepsilon}\in C^1(\RR_t;BL^1_0)$ and
$\partial_t\Phi_{\varepsilon}\in C^1(\RR_t;L^2)$. Then $t\mapsto
E(\Phi_{\varepsilon},t)$ is $C^1$ and we have
$$
\frac{d}{dt}E(\Phi_{\varepsilon},t)=2\Re\left<\left[\partial_t^2-\Delta_{\mathbf{x},z}+\frac{\mu}{z^2}\right]\Phi_{\varepsilon};\overline{\partial_t\Phi_{\varepsilon}}\right>_{H^{-1},H^1_0}=0.
$$
We get $E(\Phi_{\varepsilon},t)=E(\Phi_{\varepsilon},0)$. Since
$\Phi_{\varepsilon}\rightarrow\Phi$ in $C^0(\RR_t;BL^1_0)$ and
$\partial_t\Phi_{\varepsilon}\rightarrow\partial_t\Phi$ in
$C^0(\RR_t;L^2)$ as $\varepsilon\rightarrow 0$, we conclude that
(\ref{cons}) is satisfied by $\Phi$. Finally we establish the result of finite velocity propagation by the usual way. For a sake of simplicity, we assume the field is real valued. Given $R_1,\;R_2>0$, $T>0$, $\mathbf{x}_0\in\RR^3$, and $z_0\in\RR$ such that $\mid \mathbf{x}_0\mid>R_1+T$, $\mid z_0\mid>R_2+T$, we integrate the Pointing vector
$$
\overrightarrow{p}(t,\mathbf{x},z):=\left(
\begin{array}{c}
\mid\nabla_{t,{\mathbf x}}\Phi(t,{\mathbf
  x},z)\mid^2+\left\vert\partial_z\Phi(t,{\mathbf
  x},z)+\frac{\alpha_{\pm}}{z}\Phi(t,{\mathbf
  x},z)\right\vert^2\\
-2\partial_t \Phi(t,{\mathbf
  x},z)\nabla_{{\mathbf x}}\Phi(t,{\mathbf
  x},z)\\
-2\partial_t \Phi(t,{\mathbf
  x},z)\left(\partial_z \Phi(t,{\mathbf
  x},z)+\frac{\alpha_{\pm}}{z}\Phi(t,{\mathbf
  x},z)\right)
\end{array}
\right)
$$
on the domain $\left\{(t,\mathbf{x},z);\;\;0\leq t\leq T,\;\;\mid\mathbf{x}-\mathbf{x}_0\mid\leq R_1+T-t,\;\;\mid z-z_0\mid\leq R_2+T-t\right\}$.
Since $\nabla_{t,\mathbf{x},z}\cdotp \overrightarrow{p}=0$, we get the control of the local energy :
\begin{equation*}
\begin{split}
\int_{\mid \mathbf{x}-\mathbf{x}_0\mid\leq R_1}&\int_{\mid z-z_0\mid\leq R_2}\mid\nabla_{t,{\mathbf x}}\Phi(T,{\mathbf
  x},z)\mid^2+\left\vert\partial_z\Phi(T,{\mathbf
  x},z)+\frac{\alpha_{\pm}}{z}\Phi(T,{\mathbf
  x},z)\right\vert^2d{\mathbf
  x}dz\\
&\leq
\int_{\mid \mathbf{x}-\mathbf{x}_0\mid\leq R_1+T}\int_{\mid z-z_0\mid\leq R_2+T}\mid\nabla_{t,{\mathbf x}}\Phi(0,{\mathbf
  x},z)\mid^2+\left\vert\partial_z\Phi(0,{\mathbf
  x},z)+\frac{\alpha_{\pm}}{z}\Phi(0,{\mathbf
  x},z)\right\vert^2d{\mathbf
  x}dz,
\end{split}
\end{equation*}
and the result of finite velocity propagation is a straightforward consequence of this estimate.
\fin

In brane cosmology it is important to express the fields propagating in the Anti-de Sitter universe, as a superposition of particular solutions with an infinite energy, called the Kaluza-Klein tower, by decoupling the variables $(t,\mathrm{x})$ and $z$,  the space variable of depth.
\begin{Theorem}
For any
  $\Phi_0\in BL_0^1(\RR^3\times]0,\infty[)$, $\Phi_1\in L^2(\RR^3\times]0,\infty[)$, the solution $\Phi$ of (\ref{eq}), (\ref{reg})  and (\ref{ci}) can be expressed as
\begin{equation}
\begin{split}
\Phi(t,{\mathbf
  x},z)&=\lim_{M\rightarrow\infty}\int_0^{M}\phi_m(t,{\mathbf x})\sqrt{mz}J_{\lambda}(mz)dm\;\;\; in\;\;\;C^0\left(\RR_t;BL^1_0\left(\RR^3\times]0,\infty[\right)\right),\\
\partial_t\Phi(t,{\mathbf
  x},z)&=\lim_{M\rightarrow\infty}\int_0^{M}\partial_t\phi_m(t,{\mathbf x})\sqrt{mz}J_{\lambda}(mz)dm\;\;\;in\;\;\;C^0\left(\RR_t;L^2\left(\RR^3\times]0,\infty[\right)\right),
\end{split}
    \label{decompo}
  \end{equation}
where
\begin{equation}
\lambda:=\sqrt{\mu+\frac{1}{4}},
 \label{nu}
\end{equation}
and for any $T>0$
 \begin{equation}
\begin{split}
&\phi_m\in L^{2}\left(]0,\infty[_m;C^0\left([-T,T]_t;BL^1\left(\RR^3_{\mathbf x}\right)\right)\right)\cap
L^{2}_{loc}\left(]0,\infty[_m;C^0\left([-T,T]_t;H^1\left(\RR^3_{\mathbf x}\right)\right)\right),\\
&\partial_t\phi_m\in L^{2}\left(]0,\infty[_m;C^0\left([-T,T]_t;L^2\left(\RR^3_{\mathbf x}\right)\right)\right)
\end{split}
\label{fim}
\end{equation}
 is solution for almost all $m>0$,
  of 
\begin{equation}
\partial^2_t\phi_{m}-\Delta_{\mathbf
  x}\phi_{m}+m^2\phi_m=0,\;\;t\in\RR,\;\;\mathbf{x}\in\RR^3.
\label{kgfim}
\end{equation}
 Moreover,
\begin{equation}
\parallel \Phi_0\parallel_{BL^1_0}^2+
\parallel \Phi_1\parallel_{L^2}^2=
\int_0^{\infty}\parallel\nabla_{t,\mathbf{x}}\phi_m(t)\parallel^2_{L^2(\RR^3)}
+m^2\parallel\phi_m(t)\parallel^2_{L^2(\RR^3)}dm.
\label{deconer}
\end{equation}
\label{theosp} 
\end{Theorem}

{\it Proof of Theorem \ref{theosp}.}
We shall use some results on the Sturm-Liouville theory (see {\it e.g.} \cite{naimark}, \cite{titchmarsh},  \cite{titchmarsh1}). Given $\mu>-\frac{1}{4}$, we consider the Bessel operator
\begin{equation}
\mathrm{P}_{\mu}:=-\frac{d^2}{dz^2}+\frac{\mu}{z^2},
 \label{}
\end{equation}
and we put
\begin{equation}
\mathfrak{D}(\mu):=\left\{u\in
  L^2(]0,\infty[);\;\mathrm{P}_{\mu}u\in L^2(]0,\infty[)\right\}.
 \label{}
\end{equation}
We introduce the densely defined operator ${\mathbf h}_{\mu}$ on $ L^2(]0,\infty[)$ defined by
\begin{equation}
Dom({\mathbf h}_{\mu})=\mathfrak{D}_0(\mu):=\mathfrak{D}(\mu)\cap H^1_0(]0,\infty[),\;\;\forall u\in Dom({\mathbf h}_{\mu})\;\;{\mathbf h}_{\mu} u=\mathrm{P}_{\mu}u.
  \label{hh}
\end{equation}
The Hardy inequality assures that
\begin{equation}
\mathfrak{D}_0(\mu)=\left\{u\in \mathfrak{D}(\mu);\,\;u',\, z^{-1}u\in
  L^2(]0,\infty[)\right\}.
 \label{}
\end{equation}
It is easy to prove that $\mathbf{h}_{\mu}$ is a positive symmetric operator on $L^2(]0,\infty[)$ and the variational method shows that
$Ran(\mathbf{h}_{\mu}+1)=L^2(]0,\infty[)$, hence  $\mathbf{h}_{\mu}$ is self-adjoint. In fact, according to \cite{kalf}, \cite{rosenberger},  $\mathbf{h}_{\mu}$ is just the Friedrichs extension of the differential operator $\mathrm{P}_{\mu}$ and 
\begin{equation}
\forall u\in \mathfrak{D}_0(\mu),\;\;\lim_{z\rightarrow 0^+}z^{-\frac{1}{2}+\lambda}u(z)=\lim_{z\rightarrow 0^+}z^{\frac{1}{2}+\lambda}u'(z) =0.
 \label{uzoo}
\end{equation}

When $\mu\geq\frac{3}{2}$ ($i.e.$ $\lambda\geq 1$), $\mathrm{P}_{\mu}$ is essentially self-adjoint on $C^{\infty}_0(]0,\infty[)$ and $
\mathfrak{D}_0(\mu)=\mathfrak{D}(\mu)$.
When $\mu\in]-\frac{1}{4}, \frac{3}{4}[$ ($i.e.$  $\lambda\in]0,1[$), $\mathrm{P}_{\mu}$ is in the limit-circle case at the origin, singular for $\mu\neq 0$, and all the self-adjoint extensions $\mathbf{h}^{\omega}$, are characterized by a boundary condition at $z=0$ associated to any $\omega\in\mathfrak{D}\setminus\mathfrak{D}_0$ : $Dom(\mathbf{h}^{\omega})=\left\{u\in\mathfrak{D};\,\lim_{z\rightarrow 0^+}u'\overline{\omega}-u\overline{\omega'}=0\right\}$. As regards the Friedrichs extension $\mathbf{h}_{\mu}$,  the following sharpened asymptotics are established in \cite{everitt} for $u\in Dom(\mathbf{h}_{\mu})$ :
\begin{equation}
\lim_{z\rightarrow 0^+}z^{-\frac{1}{2}-\lambda}u(z)=K_{\lambda}(u):=u(1)+\int_0^1t^{-2\lambda-1}\left(\int_0^tz^{\lambda +\frac{1}{2}}\mathrm{P}_{\mu}u(z)dz\right)dt,
 \label{uzo}
\end{equation}
\begin{equation}
\lim_{z\rightarrow 0^+}z^{\frac{1}{2}-\lambda}u'(z)=\left(\lambda+\frac{1}{2}\right)K_{\lambda}(u),\;\;\lim_{z\rightarrow 0^+}u(z)u'(z)=0.
 \label{u'zo}
\end{equation}

The spectral representation is given by the Hankel transform of $u\in L^2(]0,\infty[)$,
\begin{equation}
\mathrm{H}_{\lambda}u(m):=\lim_{R\rightarrow\infty} \hat{H}^R_{\lambda}u\;\;in\;\;L^2\left(]0,\infty[_m,dm\right),\;\;\hat{H}^R_{\lambda}u(m):=\int_0^R\sqrt{mz}J_{\lambda}(mz)u(z)dz.
 \label{hankel}
\end{equation}
It is a well known result that for any $\lambda>0$, $\mathrm{H}_{\lambda}$ is an involutive isometry from $L^2\left(]0,\infty[_z,dz\right)$ onto  $L^2\left(]0,\infty[_m,dm\right)$ (\cite{titchmarsh1}, Theorem 129 with $\Re(s)=2^{s-\frac{1}{2}}\frac{\Gamma\left(\frac{1}{2}\lambda+\frac{1}{2}s+\frac{1}{4}\right)}{\Gamma\left(\frac{1}{2}\lambda-\frac{1}{2}s+\frac{1}{4}\right)}$, $\lambda>-1$) :
\begin{equation}
u(z)=\lim_{R\rightarrow\infty} \check{H}^R_{\lambda}\left[\mathrm{H}_{\lambda}u\right](z)\;\;in\;\;L^2\left(]0,\infty[_z,dz\right),\;\;\check{H}^R_{\lambda}v(z):=\int_0^R\sqrt{mz}J_{\lambda}(mz)v(m)dm.
 \label{hankrec}
\end{equation}
More generally, for any $u\in L^2\left(\RR^3_{\mathrm{x}}\times]0,\infty[_z\right)$ the Fubini theorem implies that
$$
\parallel u-\check{H}^R_{\lambda}\left[\mathrm{H}_{\lambda}u\right]\parallel^2_{L^2\left(\RR^3_{\mathrm x}\times]0,\infty[_z\right)}=
\int_{\RR^3}\parallel u(\mathrm{x},.)-\check{H}^R_{\lambda}\left[\mathrm{H}_{\lambda}u(\mathrm{x},.)\right]\parallel^2_{L^2\left(]0,\infty[_z\right)}d\mathrm{x},
$$
and since $\parallel u(\mathrm{x},.)-\check{H}^R_{\lambda}\left[\mathrm{H}_{\lambda}u(\mathrm{x},.)\right]\parallel^2_{L^2\left(]0,\infty[_z\right)}\leq 4 \parallel u(\mathrm{x},.) \parallel^2_{L^2\left(]0,\infty[_z\right)}$, we deduce from (\ref{hankrec}) an the dominated convergence theorem that
$$
u(\mathrm{x},z)=\lim_{R\rightarrow\infty} \check{H}^R_{\lambda}\left[\mathrm{H}_{\lambda}u(\mathrm{x},.)\right](z)\;\;in\;\;L^2\left(\RR^3_{\mathrm x}\times]0,\infty[_z\right).
 $$
Furthermore, since $ \check{H}^R_{\lambda}$ and $\mathrm{H}_{\lambda}$ are isometric, this limit is uniform on the compacts of $ L^2\left(\RR^3_{\mathrm{x}}\times]0,\infty[_z\right)$, hence we conclude that for any $u\in C^0\left(\RR_t;L^2\left(\RR^3_{\mathrm{x}}\times]0,\infty[_z\right)\right)$ we have
\begin{equation}
u(t,\mathrm{x},z)=\lim_{R\rightarrow\infty} \check{H}^R_{\lambda}\left[\mathrm{H}_{\lambda}u(t,\mathrm{x},.)\right](z)\;\;in\;\;C^0\left(\RR_t;L^2\left(\RR^3_{\mathrm x}\times]0,\infty[_z\right)\right).
 \label{}
\end{equation}
Finally for $u\in Dom(\mathbf{h}_{\mu})$ we have for all $m>0$
\begin{equation}
\mathrm{H}_{\lambda}\mathbf{h}_{\mu}u(m)=m^2\mathrm{H}_{\lambda}u(m).
 \label{}
\end{equation}
This property is easily obtained with an integration by part by using the asymptotic behaviours  at the origin (\ref{uzo}), (\ref{u'zo}). and the equality
\begin{equation}
\mathrm{P}_{\mu}\left(\sqrt{mz}J_{\lambda}(mz)\right)=m^2\sqrt{mz}J_{\lambda}(mz),\;\;0<m,\;\;0<z.
 \label{}
\end{equation}

We are now ready to apply these properties to prove the theorem. Given $t\in\RR$, for almost all $\mathbf{x}\in\RR^3$, the map
$z\mapsto\nabla_{\mathbf{x}}\Phi(t,\mathbf{x},z)$  belongs to
$L^2(\RR^+_z)$. Thus  for almost all $m>0$ we can  introduce
\begin{equation*}
\phi_m(t,\mathbf{x}):=\mathrm{H}_{\lambda}\Phi(t,\mathbf{x},.)(m),
\end{equation*}
that belongs to  $L^2\left(\RR^+_m;BL^1(\RR^3_{\mathbf x})\right)$. Moreover, 
since $\Phi\in C^0\left(\RR_t;L^2\left(\RR^+_z;BL^1\left(\RR^3_{\mathbf x}\right)\right)\right)$
  with $\partial_t\Phi\in
  C^0\left(\RR_t;L^2\left(]0,\infty[_z\times\RR^3\right)\right)$, then
  \begin{equation}
\phi_m\in
 C^0\left(\RR_t;L^2\left(\RR^+_m;BL^1\left(\RR^3_{\mathbf x}\right)\right)\right),
 \;\; \partial_t\phi_m\in
  C^0\left(\RR_t;L^2\left(\RR^+_m;L^2\left(\RR^3_{\mathbf
        x}\right)\right)\right),
    \label{rgfipl}
  \end{equation}
 therefore (\ref{decompo}) is established.
We also have :
  \begin{equation}
\left\{
\begin{split}
\phi_m(0,\mathbf{x})=\mathrm{H}_{\lambda}\left(\Phi_0(\mathbf{x},.)\right)(m)\in L^2\left(\RR^+_m;BL^1\left(\RR^3_{\mathbf x}\right)\right),
\\
\partial_t\phi_m(0,\mathbf{x})=  \mathrm{H}_{\lambda}\left(\Phi_1(\mathbf{x},.)\right)(m)  \in L^2\left(\RR^+_m;L^2\left(\RR^3_{\mathbf x}\right)\right).
\end{split}
\right.
    \label{cifit}
  \end{equation}
Moreover we have 
\begin{equation*}
\nabla_{t,\mathbf{x}}\Phi(t,\mathbf{x},z)=\mathrm{H}_{\lambda}\left[\nabla_{t,\mathbf{x}}\phi_m(t,\mathbf{x})\right](z)\in C^0\left(\RR_t;L^2(\RR^3_{\mathbf{x}}\times]0,\infty[_z)\right),
\end{equation*}
hence
\begin{equation}
\int_0^{\infty}\int_{\RR^3_{\mathbf
    x}}\mid\nabla_{t,\mathbf{x}}\Phi(t,\mathbf{x},z)\mid^2
dzd\mathbf{x}=\int_0^{\infty}\int_{\RR^3_{\mathbf
    x}}\mid\nabla_{t,\mathbf{x}}\phi_m(t,\mathbf{x},m)\mid^2
dmd\mathbf{x}.
\label{nablafi}
\end{equation}
We remark that for $u\in{\mathfrak D}(\mathbf{h}_{\mu})$ we have 
$$
\int_0^{\infty}\left\vert u'(z)+\frac{\alpha_{\pm}}{z}u(z)\right\vert^2dz=
<\mathbf{h}_{\mu}u,u>_{L^2(\RR^+)}=\int_0^{\infty}\left\vert \mathrm{H}_{\lambda}u(m)\right\vert^2m^2dm.
$$
This equality can be extended by density into an isometry from  the
closure of ${\mathfrak D}(\mathbf{h}_{\mu})$ for the norm associated
to the first integral, onto $L^2(\RR^+_m, m^2dm)$. We deduce that
$m\phi_m\in
  C^0\left(\RR_t;L^2\left(\RR^+_m;L^2\left(\RR^3_{\mathbf
        x}\right)\right)\right)$ hence with (\ref{rgfipl}) 
  \begin{equation}
\forall a>0,\;\;\phi_m\in
 C^0\left(\RR_t;L^2\left([a,\infty[_m;H^1\left(\RR^3_{\mathbf x}\right)\right)\right),
    \label{regfih}
  \end{equation}
and 
 \begin{equation}
\int_0^{\infty}\int_{\RR^3_{\mathbf x}}\left\vert
    \partial_z\Phi(t,\mathbf{x},z)+\frac{\alpha_{\pm}}{z}\Phi(t,\mathbf{x},z)\right\vert^2dzd\mathbf{x}=
\int_0^{\infty}\int_{\RR^3_{\mathbf
    x}}m^2\mid\phi_m(t,\mathbf{x},m)\mid^2
dmd\mathbf{x}.
\label{dzfi}
  \end{equation}
Now (\ref{deconer}) follows from (\ref{nablafi}) and (\ref{dzfi}).\\

Now we establish that $\phi_m$ is a finite energy solution of the
Klein-Gordon equation for almost all $m>0$. Thanks to (\ref{deconer}),
we see that the map $(\Phi_0,\Phi_1)\mapsto (\phi_m,\partial_t\phi_m)$ is
continuous from $BL_0^1(\RR^3\times]0,\infty[)\times L^2(\RR^3\times]0,\infty[)$ to
$C^0\left(\RR_t;L^2\left(\RR^+_m;BL^1(\RR^3_{\mathbf x})\right)\right)\times C^0\left(\RR_t;L^2\left(\RR^+_m;L^2(\RR^3_{\mathbf x})\right)\right)$, hence it is sufficient to prove
that  
\begin{equation}
(\partial_t^2-\Delta_{\mathbf x}+m^2)\phi_m=0\;\;in\;\;{\mathcal D}'\left(\RR_t\times\RR^3_{\mathbf x}\times]0,\infty[_m\right)
  \label{kgm}
\end{equation}
 for a
dense set of initial data. We choose $\Phi_0,\;\Phi_1\in
C^{\infty}_0(\RR^3\times]0,\infty[)$. Then the solution $\Phi$ of the Cauchy
problem is compactly supported in space at each time and since $\Delta_{\mathbf x}$ and $-\Delta_{{\mathbf x},z}+\frac{\mu}{z^2}$
are commuting, $\Delta_{\mathbf{x}}\Phi$ is also a finite energy solution, and we have :
\begin{equation*}
\Phi,\;\;\Delta_{\mathbf{x}}\Phi\in C^2\left(\RR_t;L^2\left(\RR^3\times]0,\infty[\right)\right)\cap C^1\left(\RR_t;H^1_0\left(\RR^3\times]0,\infty[\right)\right).
\end{equation*}
This implies that 
$$
\left[-\partial_z^2+\frac{\mu}{z^2}\right]\Phi\in  C^0\left(\RR_t;L^2(\RR^3\times]0,\infty[)\right)
$$
We deduce that 
$$
\Phi\in C^0\left(\RR_t;L^2\left(\RR^3_{\mathbf{x}};\mathfrak{D}_0(\mu)\right)\right),
$$
therefore $\phi_m$belongs to $C^1\left(\RR_t;L^2\left(\RR^+_m;H^1(\RR^3_{\mathbf x})\right)\right)\times C^2\left(\RR_t;L^2\left(\RR^+_m;L^2(\RR^3_{\mathbf x})\right)\right)$, and
\begin{equation*}
(\partial_t^2-\Delta_{\mathbf
  x}+m^2)\phi_m=\mathrm{H}_{\lambda}\left(\partial_t^2\Phi-\Delta_{{\mathbf
  x},z}\Phi-\frac{\mu}{z^2}\Phi\right)=0.
\end{equation*}

It remains to prove that in the general case where
$(\Phi_0,\Phi_1)\in BL^1_0\left(\RR^3\times]0,\infty[\right)\times L^2\left(\RR^3\times]0,\infty[\right)$, $\phi_m$ belongs to $C^0\left(\RR_t;H^1\left(\RR^3_{\mathbf x}\right)\right)\cap
  C^1\left(\RR_t;L^2\left(\RR^3_{\mathbf x}\right)\right)$ for almost
  all $m>0$. We have established that for $0<a$, $\phi_m\in C^0\left(\RR_t;L^2\left([a,\infty[_m;H^1\left(\RR^3_{\mathbf x}\right)\right)\right)\cap
  C^1\left(\RR_t;L^2\left([a,\infty[_m;L^2\left(\RR^3_{\mathbf
          x}\right)\right)\right)$ is solution of 
  (\ref{cifit}) and (\ref{kgm}). We have proved in \cite{RS}, p. 829-830, that this Cauchy problem is
  well posed in this functional framework and the solution belongs to $ C^0\left(\RR_t;H^1\left(\RR^3_{\mathbf x}\right)\right)\cap
  C^1\left(\RR_t;L^2\left(\RR^3_{\mathbf
          x}\right)\right)$ for almost all $m>0$.
  
\fin


Since the smooth solutions of the massive Klein-Gordon equation on the 3+1 dimensional Minkowski space-time decay as $\mid t\mid^{-\frac{3}{2}}$, we can use lemma 4.3 of \cite{RS} to obtain the same rate of decay uniformly for a suitable class of solutions of (\ref{eq}). The result above states a $L^1-L^{\infty}$ estimate of von Wahl type in weighted spaces, in particular the factor $z^{-\lambda-\frac{1}{2}}$ in the uniform bound expresses that the horizon is repulsive. In the next part we establish some more strong properties for certain values of the mass, in particular for the gravitational or electromagnetic fluctuations.

\begin{Theorem}
  \label{propdecay}
There exists $C>0$ such that any finite energy solution $\Phi$ of (\ref{eq}), (\ref{ci}) satisfies the following estimate with $\lambda=\sqrt{\mu+\frac{1}{4}}$, provided the $L^1$ norms in the right member are finite :
\begin{equation}
\begin{split}
\left\lVert z^{-\lambda-\frac{1}{2}}\Phi(t,.)\right\rVert_{L^{\infty}(\RR^3_{\mathrm x}\times]0,\infty[_z)}&\leq \\
C \mid t\mid^{-\frac{3}{2}}\sum_{j=0,1}\sum_{\mid\alpha\mid+j\leq 3}&\left\|\partial^{\alpha}_{\mathrm x}\Phi_j\right\rVert_{L^1(\RR^3_{\mathrm x}\times]0,\infty[_z)}
+\left\|\partial^{\alpha}_{\mathrm x}\left(-\partial_z^2+\frac{\mu}{z^2}\right)^{\left[\frac{\lambda+3-\mid\alpha\mid-j}{2}\right]+1}\Phi_j\right\|_{L^1(\RR^3_{\mathrm x}\times]0,\infty[_z)}.
\end{split}
 \label{decayLun}
\end{equation}
\end{Theorem}

{\it Proof of Theorem \ref{propdecay}.} It sufficient to consider the case $\Phi_j\in C^{\infty}_0\left(\RR^3_{\mathrm x}\times]0,\infty[_z\right)$. Since the Bessel function satisfies $\mid J_{\lambda}(x)\mid \leq C x^{\lambda}$, we can write
$$
\int_0^M\left\vert\sqrt{mz}J_{\lambda}(mz)\phi_m(t,\mathrm{x})\right\vert dm\leq Cz^{\lambda+\frac{1}{2}}\int_0^Mm^{\lambda+\frac{1}{2}}\parallel \phi_m(t,.)\parallel_{L^{\infty}(\RR^3_{\mathrm x})}dm.
$$
We have $$\phi_m(0,\mathrm{x})=\phi_m^0(\mathrm{x}):=\int_0^{\infty}\sqrt{mz}J_{\lambda}(mz)\Phi_0(\mathrm{x},z)dz,$$
$$\partial_t\phi_m(0,\mathrm{x})=\phi_m^1(\mathrm{x}):=\int_0^{\infty}\sqrt{mz}J_{\lambda}(mz)\Phi_1(\mathrm{x},z)dz.$$
We remark that for all $k\in\NN$
$$
\phi_m^j(\mathrm{x})=m^{-2k}\int_0^{\infty}\sqrt{mz}J_{\lambda}(mz) \left(-\partial_z^2+\frac{\mu}{z^2}\right)^k\Phi_j(\mathrm{x},z)dz.
$$
Since $\sqrt{x}J_{\lambda}(x)\in L^{\infty}(]0,\infty[)$, we get
$$
\parallel \partial_{\mathrm{x}}^{\alpha}\phi_m^j)\parallel_{L^1(\RR^3_{\mathrm x})}\leq C(1+m)^{-2k}\parallel\Phi_j\parallel_{\alpha,k}$$
with
$$
\parallel\Phi_j\parallel_{\alpha,k}:=\left\|\partial^{\alpha}_{\mathrm x}\Phi_j\right\rVert_{L^1(\RR^3_{\mathrm x}\times]0,\infty[_z)}
+\left\|\partial^{\alpha}_{\mathrm x}\left(-\partial_z^2+\frac{\mu}{z^2}\right)^k\Phi_j\right\|_{L^1(\RR^3_{\mathrm x}\times]0,\infty[_z)}.
$$
The equation (4.4) in \cite{RS} assures that
$$
\parallel \phi_m(t,.)\parallel_{L^{\infty}(\RR^3_{\mathrm x})}\leq C(1+m\mid t\mid)^{-\frac{3}{2}}\sum_{j=0,1}\sum_{\mid\alpha\mid+j\leq 3}m^{3-\mid\alpha\mid-j}\parallel\partial_{\mathrm x}^{\alpha}\phi_m^j\parallel_{L^1(\RR^3_{\mathrm x})},
$$
Then we get
\begin{equation*}
\begin{split}
\int_0^M\left\vert\sqrt{mz}J_{\lambda}(mz)\phi_m(t,\mathrm{x})\right\vert& dm\leq \\
&Cz^{\lambda+\frac{1}{2}}\sum_{j=0,1}\sum_{\mid\alpha\mid+j\leq 3}\int_0^Mm^{\lambda+\frac{1}{2}+3-\mid\alpha\mid-j}(1+m)^{-2k}(1+m\mid t\mid)^{-\frac{3}{2}}dm\parallel\Phi_j\parallel_{\alpha,k}\\
&Cz^{\lambda+\frac{1}{2}}\mid t\mid^{-\frac{3}{2}}\sum_{j=0,1}\sum_{\mid\alpha\mid+j\leq 3}\int_0^Mm^{\lambda+2-\mid\alpha\mid-j}(1+m)^{-2k}dm\parallel\Phi_j\parallel_{\alpha,k}.
\end{split}
\end{equation*}
Since $\lambda>0$, it is sufficient to choose $k=\left[\frac{\lambda+3-\mid\alpha\mid-j}{2}\right]+1$ to assure that
$$
\int_0^{\infty}m^{\lambda+2-\mid\alpha\mid-j}(1+m)^{-2k}dm<\infty.
$$
We conclude that when the right member of (\ref{decayLun}) is finite, given $t\in\RR$, the limit (\ref{decompo}) holds in $L^{\infty}\left(\RR^3_{\mathrm x}\times]0,\infty[_z;z^{-\lambda-\frac{1}{2}}d\mathrm{x}dz\right)$ and its norm is estimated by  (\ref{decayLun}).
\fin

\section{The case of the mass $\mu=\frac{\nu^2-1}{4}$, $\nu\in\NN^*$.}

The equations for the gravitational fluctuations ($\mu=\frac{15}{4}$) or the electromagnetic perturbations ($\mu=\frac{3}{4}$) belong to a large class for which $\mu=\frac{\nu^2-1}{4}$, $\nu\in\NN^*$. For these values of $\mu$, the finite energy solutions are closely linked to the finite energy solutions of the free wave equation on the Minkowski space-time with a higher dimension. The method rests on
a very simple observation : we can consider the fifth space-like dimension $z>0$, as the radial
coordinate of some euclidean high-dimensional space $\RR^N_{\mathbf z}$, $N\geq 2$,
i.e. $z=\mid \mathbf{z}\mid$. We denote $Y_{l,m}$ the generalized
spherical harmonics that form a orthonormal basis of $L^2(S^{N-1})$ of
eigenfunctions of the Laplace-Beltrami operator satisfying $\Delta_{S^{N-1}}Y_{l,m}=-l(l+N-2)Y_{l,m}$. Here $l,m\in \NN$ and $m$ is bounded by the dimension of  the space of harmonic  homogeneous polynomials of degree $l$ in $ N$ variables. Then it is straightforward to check that $\Phi\in
L^1_{loc}(\RR_t\times\RR^3_{\mathbf x}\times]0,\infty[_z)$ is solution
of (\ref{eq}) with
\begin{equation}
\mu=-\frac{1}{4}+\left(\frac{N}{2}+l-1\right)^2,\;\;l,N\in\NN,\;\;0\leq
l,\;\;2\leq
N,
  \label{mukg}
\end{equation}
if and only if
\begin{equation}
\Psi(t,\mathbf{x},\mathbf{z}):=\mid
\mathbf{z}\mid^{-\frac{N-1}{2}}\Phi(t,\mathbf{x},\mid
\mathbf{z}\mid)Y_{l,m}\left(\frac{\mathbf{z}}{\mid \mathbf{z}\mid}\right)
  \label{psi}
\end{equation}
is a solution in
$L^1_{loc}\left(\RR_t\times\RR^3_{\mathbf{x}}\times\left(\RR^N_{\mathbf{z}}\setminus\{0\}\right)\right)$
  of

  \begin{equation}
\left(\partial_t^2-\Delta_{\mathbf{x}}-\Delta_{\mathbf
    z}\right)\Psi=0,\;\;(t,{\mathbf x},{\mathbf z})\in\RR\times\RR^3\times\left(\RR^N\setminus\{0\}\right).
    \label{}
  \end{equation}
Now the crucial point is that when $\Phi$ is a finite energy solution
in $AdS^5$,
then $\Psi$ is a finite energy free wave in the whole Minkowski space-time
$\RR_t\times\RR^{3+N}_{\mathbf{x},\mathbf{z}}$.

\begin{Lemma}
Let $\Phi$ be satisfying (\ref{reg}), a solution of (\ref{eq}) where
$\mu$ is given by (\ref{mukg}) for some integers $N\geq 3$ and $l\geq
0$, or $N=2$ and $l\geq 1$. Then $\Psi$ defined by (\ref{psi}) satisfies
\begin{equation}
\Psi\in C^0\left(\RR_t;BL^1\left(\RR^{3+N}_{\mathbf{x},\mathbf{z}}\right)\right),\;\;
\partial_t\Psi\in C^0\left(\RR_t;L^2\left(\RR^{3+N}_{\mathbf{x},\mathbf{z}}\right)\right),
  \label{regol}
\end{equation}
\begin{equation}
\partial_t^2\Psi-\Delta_{\mathbf{x},\mathbf{z}}\Psi=0,\;\;(t,{\mathbf x},{\mathbf z})\in\RR\times\RR^{3+N}.
  \label{eqpsi}
\end{equation}
  \label{lembo}
\end{Lemma}

{\it Proof.}
We introduce the operator
\begin{equation}
\begin{split}
\Pi_{l,m}:\; C^{\infty}_0\left(\RR^3_{\mathbf
    x}\times]0,\infty[_z\right)&\longrightarrow 
C^{\infty}_0\left(\RR^3_{\mathbf
    x}\times\left(\RR^N_{\mathbf z}\setminus\{0\}\right)\right),\\
\phi(\mathbf{x},z)&\longmapsto
\left(\Pi_{l,m}\phi\right)(\mathbf{x},\mathbf{z}):=\mid\mathbf{z}\mid^{-\frac{N-1}{2}}\phi(\mathbf{x},\mid\mathbf{z}\mid)Y_{l,m}\left(\frac{\mathbf
  z}{\mid\mathbf{z}\mid}\right).
\end{split}
\label{pilm}
\end{equation}
It is clear that $\Pi_{l,m}$ can be extended in an isometry from $L^2\left(\RR^3_{\mathbf
    x}\times]0,\infty[_z\right)$ to
$L^2\left(\RR^{3+N}_{\mathbf{x},\mathbf{z}}\right)$. Moreover we have
$$
\Delta_{\mathbf{x},\mathbf{z}}\Pi_{l,m}=\Pi_{l,m}\left(\Delta_{\mathbf{x}}+\partial_z^2-\frac{\mu}{z^2}\right),
$$
therefore given $\phi,\;\phi'\in  C^{\infty}_0\left(\RR^3_{\mathbf
    x}\times]0,\infty[_z\right)$, we have
\begin{equation}
\begin{split}
\left<\Pi_{l,m}\phi,\Pi'_{l',m'}\phi'\right>_{BL^1(\RR^{3+N})}&=
-<\Delta_{\mathbf{x},\mathbf{z}}\Pi_{l,m}\phi,\Pi_{l',m'}\phi'>_{L^2(\RR^{3+N})}\\
&=
-<\Pi_{l,m}\left(\Delta_{\mathbf{x}}+\partial_z^2-\frac{\mu}{z^2}\right)\phi,\Pi_{l',m'}\phi'>_{L^2(\RR^{3+N})}\\
&=
-<\left(\Delta_{\mathbf{x}}+\partial_z^2-\frac{\mu}{z^2}\right)\phi,\phi'>_{L^2(\RR^3\times]0,\infty[)}\delta_{l,l'}\delta_{m,m'}\\
&=\left<\phi,\phi'\right>_{BL^1_0(\RR^3\times]0,\infty[) }\delta_{l,l'}\delta_{m,m'},
\end{split}
\label{fifiprime}
\end{equation}
where $\delta_{a,b}$ is the symbol of Kronecker.
We conclude that $\Pi_{l,m}$ is a continuous linear map from
$BL^1_0(\RR^3\times]0,\infty[)$ to $BL^1(\RR^{3+N})$, and also that the
map (\ref{psi}) is continuous from
$C^0(\RR_t;BL^1_0(\RR^3\times]0,\infty[))$ to
$C^0(\RR_t;BL^1(\RR^{3+N}))$, and from $C^0(\RR_t;L^2(\RR^3\times]0,\infty[))$ to
$C^0(\RR_t;L^2(\RR^{3+N}))$. Hence it is sufficient to establish that $\Psi:=\Pi_{l,m}\Phi$ satisfies (\ref{regol}) and (\ref{eqpsi})
when $\Phi_0=\Phi(0,.)$ and $\Phi_1=\partial_t\Phi(0,.)$ belong to
$C^{\infty}_0\left(\RR^3_{\mathbf x}\times]0,\infty[_z\right)$. For
such a solution, we consider the solution $\hat\Phi$ of the Cauchy problem
for the wave equation in the $1+3+N$ dimensional Minkowski space :
$$
\partial^2_t\hat\Phi-\Delta_{\mathbf{x},\mathbf{z}}\hat\Phi=0\;\;in\;\;\RR^{1+3+N}_{t,\mathbf{x},\mathbf{z}},\;\;\hat\Phi(0)=\Pi_{l,m}\Phi_0,\;\;\partial_t\hat\Phi(0)=\Pi_{l,m}\Phi_1.
$$
We know that $\hat\Phi\in
C^{\infty}(\RR_t;C^{\infty}_0(\RR^{3+N}_{\mathbf{x},\mathbf{z}}))$ and
the proof will be achieved if we prove that $\Psi$ is equal to $\hat\Phi$.
We denote $\Pi_{l,m}^*$ the adjoint of  $\Pi_{l,m}$, defined for $\psi\in  L^2_{loc}\left(\RR^3_{\mathbf{x}}\times\RR^N_{\mathbf
z}\right)$ by
$$
\Pi_{l,m}^*\psi(\mathbf{x},z):=z^{\frac{N-1}{2}}\int_{S^{N-1}}\psi(\mathbf{x},z\pmb{\omega})\overline{Y_{l,m}({\pmb{\omega}})}d{\pmb{\omega}}.
$$
For all $\phi\in L^2(\RR^3_{\mathbf x}\times]0,\infty[_z)$ and all $\psi\in  L^2_{loc}\left(\RR^3_{\mathbf{x}}\times\RR^N_{\mathbf
z}\right)$, we have
$$
\Pi_{l,m}^*\Pi_{l,m}\phi=\phi,\;\;\psi=\sum_{l,m}\Pi_{l,m}\Pi_{l,m}^*\psi,
$$
moreover we can see from (\ref{fifiprime}) that
for $\psi\in C^{\infty}_0 \left(\RR^3_{\mathbf{x}}\times\left(\RR^N_{\mathbf
z}\setminus\{0\}\right)\right)$, we have
$$
\parallel\psi\parallel^2_{BL^1(\RR^3_{\mathrm{x}}\times\RR^N_{\mathrm{z}})}=\sum_{l,m}\parallel\Pi_{l,m}^*\psi\parallel^2_{BL^1_0(\RR^3_{\mathrm{x}}\times]0,\infty[_z)}.
$$
Furthermore if $\psi\in C^{\infty}_0 \left(\RR^3_{\mathbf{x}}\times\left(\RR^N_{\mathbf
z}\setminus\{0\}\right)\right)$ and $\theta\in C^{\infty}_0(\RR)$, $\theta(z)=1$ for all $\mid z\mid\leq 1$, we can easily check that if $N\geq 3$,  $\left(1-\theta(n\mid\mathrm{z}\mid)\psi(\mathrm{x},\mathrm{z})\right)$ tends to $\psi$ in $BL^1\left(\RR^3_{\mathrm{x}}\times\RR^N_{\mathrm{z}}\right)$ as $n\rightarrow\infty$. We conclude that  $C^{\infty}_0 \left(\RR^3_{\mathbf{x}}\times\left(\RR^N_{\mathbf
z}\setminus\{0\}\right)\right)$ is dense in  $BL^1\left(\RR^3_{\mathbf{x}}\times\RR^N_{\mathbf
z}\right)$ and $\Pi_{l,m}^*$ can be extended into a bounded linear map from $BL^1\left(\RR^3_{\mathbf{x}}\times\RR^N_{\mathbf
z}\right)$ to $BL^1_0\left(\RR^3_{\mathbf{x}}\times]0,\infty[_z\right)$. In the case $N=2$ and $l\geq 1$, $m=\pm 1$, we remark that
$$
\partial_z\Pi_{l,m}^*\psi(\mathrm{x},z)=\frac{1}{l}\sqrt{z}\int_0^{2\pi}\left(\sin\theta\partial_{Z_1}\psi(\mathrm{x},z\cos\theta,z\sin\theta)-
(\cos\theta\partial_{Z_2}\psi(\mathrm{x},z\cos\theta,z\sin\theta)\right)e^{iml\theta}d\theta,
$$
hence $\Pi_{l,m}^*\psi\in BL^1\left(\RR^3_{\mathbf{x}}\times]0,\infty[_z\right)$ and since $\Pi_{l,m}^*\psi(\mathrm{x},0)=0$,
$\Pi_{l,m}^*$ can be extended into a bounded linear map from $BL^1\left(\RR^3_{\mathbf{x}}\times\RR^2_{\mathbf
z}\right)$ to $BL^1_0\left(\RR^3_{\mathbf{x}}\times]0,\infty[_z\right)$ again.
Therefore we can put $F(t,\mathbf{x},z):=\Pi_{l,m}^*\hat\Phi(t,\mathbf{x},z)$.  We remark that
$
F\in
C^0(\RR_t;BL^1_0(\RR^{3}_{\mathbf{x}}\times]0,\infty[_z))\cap C^1(\RR_t;L^2(\RR^{3}_{\mathbf{x}}\times]0,\infty[_z))$,
$F(0,\mathbf{x},z)=\Phi_0(\mathbf{x},z)$, $\partial_tF(0,\mathbf{x},z)=\Phi_1(\mathbf{x},z)$.
Thus $F$ satisfies (\ref{reg}) and (\ref{ci}). Now we calculate
$$
\left(\partial_t^2-\Delta_{\mathbf{x}}-\partial_z^2+\frac{\mu}{z^2}\right)F(t,\mathbf{x},z)=z^{\frac{N-1}{2}}\int_{S^{N-1}}(\partial^2_t\hat\Phi-\Delta_{\mathbf{x},\mathbf{z}}\hat\Phi)(t,\mathbf{x},z{\pmb{\omega}})\overline{Y_{l,m}({\pmb{\omega}})}d{\pmb{\omega}}=0.
$$
We conclude that $F=\Phi$, and so
$$
\Psi=\Pi_{l,m}F.
$$
To end the proof we introduce the projector
$$
P_{l,m}:\;\psi\in L^2_{loc}(\RR^3_{\mathbf{x}}\times\RR^N_{\mathbf
z})\longmapsto
P_{l,m}\psi(\mathbf{x},\mathbf{z}):=\left(\int_{S^{N-1}}\psi(\mathbf{x},\mid\mathbf{z}\mid{\pmb{\omega}})\overline{Y_{l,m}({\pmb{\omega}})}d{\pmb{\omega}}\right)Y_{l,m}\left(\frac{\mathbf{z}}{\mid\mathbf{z}\mid}\right).
$$
It satisfies
$\Delta_{\mathbf{x},\mathbf{z}}P_{l,m}=P_{l,m}\Delta_{\mathbf{x},\mathbf{z}}$
on $C^{\infty}_0(\RR^{3+N})$. Thus $P_{l,m}\hat\Phi\in
C^{\infty}(\RR_t;C^{\infty}_0(\RR^{3+N}_{\mathbf{x},\mathbf{z}}))$ is solution of
$$
\partial^2_tP_{l,m}\hat\Phi-\Delta_{\mathbf{x},\mathbf{z}}P_{l,m}\hat\Phi=0\;\;in\;\;\RR^{1+3+N}_{t,\mathbf{x},\mathbf{z}},\;\;P_{l,m}\hat\Phi(0)=\Pi_{l,m}\Phi_0,\;\;\partial_tP_{l,m}\hat\Phi(0)=\Pi_{l,m}\Phi_1.
$$
By uniqueness, we deduce that $P_{l,m}\hat\Phi=\hat\Phi$. Since
$\Pi_{l,m}F=P_{l,m}\hat\Phi$, we finally get $\Psi=\hat\Phi$ and that achieves the proof.

 For the sake of completeness, we mention another way to obtain this result. We could use the expansion (\ref{decompo}) and show that $\left(\partial_t^2-\Delta_{\mathrm{x},\mathrm{z}}\right)\left[\phi_m(t,\mathrm{x})\mid\mathrm{z}\mid^{-\frac{N}{2}+1}J_{\lambda}(m\mid\mathrm{z}\mid)\right]=0$ when $\lambda=\frac{N}{2}-1$. The crucial fact is that $ -\Delta_{\mathrm{z}}\left[\mid\mathrm{z}\mid^{-\frac{N}{2}+1}J_{\lambda}(m\mid\mathrm{z}\mid)\right]=m^2 \mid\mathrm{z}\mid^{-\frac{N}{2}+1}J_{\lambda}(m\mid\mathrm{z}\mid)$ in $\RR^N_{\mathrm{z}}$.
\fin

With this result, we can easily deduce the properties of the finite
energy solutions when $\mu$ can be expressed by (\ref{mukg}), from the properties of the free waves in the
Minkowski space-time. We use just the formula
\begin{equation}
\Phi(t,\mathbf{x},z)=z^{\frac{N-1}{2}}\int_{S^{N-1}}\Psi(t,\mathbf{x},z{\pmb{\omega}})\overline{Y_{l,m}({\pmb{\omega}})}d{\pmb{\omega}},
  \label{formula}
\end{equation}
where $\Psi$ is the solution of (\ref{regol}), (\ref{eqpsi}) with
\begin{equation}
\Psi(0)=\Pi_{l,m}\Phi_0,\;\;\partial_t\Psi(0)=\Pi_{l,m}\Phi_1.
  \label{}
\end{equation}

The first statement deals with the existence of a lacuna, and also the equipartition of the energy at finite time when the
initial data is compactly supported and $\nu$ is even.

\begin{Theorem}
We assume that
\begin{equation}
\mu=n^2-\frac{1}{4},\;\;n\in\NN^*,
  \label{mu}
\end{equation}
and $\Phi_0\in BL^1_0(\RR^3\times]0,\infty[)$, $\Phi_1\in
L^2(\RR^3\times]0,\infty[)$ satisfy for some $R>0$
\begin{equation}
\mid\mathbf x\mid^2+z^2\leq R^2\Rightarrow \Phi_0(\mathbf{x},z)= \Phi_1(\mathbf{x},z)=0.
  \label{}
\end{equation}
Then the finite energy solution $\Phi$ of (\ref{eq}), (\ref{reg}) and
(\ref{ci}), satisfies
\begin{equation}
\mid\mathbf x\mid^2+z^2\leq (\mid t\mid -R)^2\Rightarrow \Phi(t,\mathbf{x},z)=0,
  \label{lacuna}
\end{equation}
and for $\mid t\mid\geq R$ the potential and kinetic energies are equal :
\begin{equation}
\int_{\RR^3}\int_0^{\infty}\mid\nabla_{{\mathbf x},z}\Phi(t,{\mathbf
  x},z)\mid^2+\frac{\mu}{z^2}\mid\Phi(t,{\mathbf x},z)\mid^2d{\mathbf x}dz=
\int_{\RR^3}\int_0^{\infty}\mid\partial_t\Phi(t,{\mathbf x},z)\mid^2d{\mathbf x}dz.
  \label{equi}
\end{equation}
  \label{h}
\end{Theorem}

 These properties are somewhat unexpected : the Huygens Principle fails for the equation (\ref{eq}) since the
number of the  space dimensions is even and the Hadamard's criterion is not satisfied (see \cite{gunther}, Theorem 1.3, p.231; we also note that (\ref{eq}) has the form of the equations considered by K.L Stellmacher that are Huygens operators iff the space dimension is odd, \cite{gunther} chapter V.4). Therefore we must not confuse the existence of this lacuna with the Huygens Principle that is a much stronger property that is not satisfied in our case. A similar situation occurs for the very simple case of the wave equation on the half line, $\partial_t^2u-\partial_z^2u=0$, $z>0$, with the Dirichlet condition $u(t,z=0)=0$. In particular, the previous result assures that there exists a lacuna for the finite energy gravitational fluctuations, since
$\mu=\frac{15}{4}=2^2-\frac{1}{4}$, and the electromagnetic fields since $\mu=\frac{3}{4}=1^2-\frac{1}{4}$. At our
knowledge, this property of the gravitational or electromagnetic waves in $AdS^5$ was not
mentioned in the literature. To a similar property of ``characteristic propagation'' in $AdS^4$, see \cite{torrence}. The equipartition of the energy at finite time is also rather surprising since it is a well-known result for the free waves just for the odd space dimension \cite{equipart}.\\

{\it Proof of Theorem \ref{h}.}
These results are direct consequences of (\ref{formula}). We take
$N=2(n+1)$ and $l=0$, and to get (\ref{lacuna}), we invoke the well known Huygens Principle
satisfied by the free waves $\Psi(t,.):=\left(\Pi_{0,0}\Phi(t,.)\right)$ in the Minkowski space-time
$\RR_t\times\RR^{3+2(n+1)}_{\mathbf{x},\mathbf{z}}$. To obtain (\ref{ener}), we use (\ref{normz}) with $\alpha=n+\frac{1}{2}$ and we check that
\begin{equation*}
\begin{split}
\int_{\RR^3}\int_0^{\infty}\mid\nabla_{{\mathbf x},z}\Phi(t,{\mathbf
  x},z)\mid^2+&\frac{\mu}{z^2}\mid\Phi(t,{\mathbf x},z)\mid^2-\mid\partial_t\Phi(t,{\mathbf x},z)\mid^2d{\mathbf x}dz\\
&=\frac{1}{\mid S^{2n+1}\mid }\int_{\RR^{3+2n+2}}\mid\nabla_{{\mathbf x},\mathbf{z}}\Psi(t,{\mathbf
  x},\mathbf{z})\mid^2-\mid\partial_t\Psi(t,{\mathbf x},\mathbf{z})\mid^2d{\mathbf x}d{\mathbf z}.
\end{split}
\end{equation*}
It is a classical result (Lax-Phillips, Duffin, see e.g. \cite{equipart}) that the last integral is zero when $\mid t\mid\geq R$, hence (\ref{equi}) follows.
\fin

Now we prove that the weak Huygens principle holds, that is to say, the singularities are propagating according to the geometrical optics, in particular, they are reflected by the horizon $z=0$ with the Descartes law. The following proposition describes the structure of the wave front set $WF(\Phi)$ of a finite energy solution. Since the principal part of the differential operator (\ref{eq}) is simply the wave equation in the flat space, we know that
$$
WF(\Phi)\subset\left\{\left(t,\mathbf{x},z;\tau,\pmb{\xi},\zeta\right)\in\RR\times\RR^3\times]0,\infty[\times\RR\times\RR^3\times\RR;\;\;\tau^2=\mid\pmb{\xi}\mid^2+\zeta^2\right\},
$$
and we are mainly concerned by the rays $\left(t+\lambda\tau,\mathbf{x}-\lambda\pmb{\xi},z-\lambda\zeta\right)_{\lambda\in\RR}\subset\RR\times\RR^3\times]0,\infty[$.

\begin{Theorem}
We assume that $\mu$ satisfies
$$
\mu=\frac{\nu^2-1}{4},\;\;\nu\in\NN^*.
$$
We consider a finite energy solution $\Phi$, and 
$$(t,\mathbf{x},z;\tau,\pmb{\xi},\zeta)\in WF(\Phi).$$
Then for any $\lambda\in\RR$ such that $\lambda\zeta\neq z$, we have
$$\left(t+\lambda\tau,\mathbf{x}-\lambda\pmb{\xi},\mid z-\lambda\zeta\mid;\tau,\pmb{\xi},\frac{z-\lambda\zeta}{\mid z-\lambda\zeta\mid}\zeta\right)\in WF(\Phi).$$
\label{propag}
\end{Theorem}

In particular, this thorem explains the role of the horizon in the propagation of the finite energy gravitational fluctuations ($\nu=4$), and the electromagnetic waves ($\nu=2$) : the boundary of the Anti-De Sitter universe is a perfect mirror. Nevertheless, the constraint on the mass is somewhat unsatisfactory, and we could expect that the time-like horizon acts like a perfectly reflecting boundary for all the fields regardless of their mass. Moreover, we can hope that the time-like horizons of the  general space-times that are asympotically Anti-de Sitter, have this same property. In a recent work, A. Vasy has proved this result for the D'Alembertian \cite{vasy}.\\

{\it Proof of Theorem \ref{propag}.} We fix $N=\nu+2$ and we define on $\RR_t\times\RR^3_{\mathbf{x}}\times\left(\RR^N_{\mathbf{z}}\setminus\{0\}\right)$, $\Psi_0(t,\mathbf{x},\mathbf{z}):=\Phi(t,\mathbf{x},\mid\mathbf{z}\mid)$. We introduce the map $f(t,\mathbf{x},\mathbf{z}):=(t,\mathbf{x},\mid\mathbf{z}\mid)$, hence $\Psi_0=\Phi\circ f:=f^*\Phi$ and the theorem on the wave front set of a pullback (see e.g. Theorem 8.2.4 in \cite{hormander1}) assures that
$$
WF(\Psi_0)\subset f^*WF(\Phi):=\left\{\left(t,\mathbf{x},\mathbf{z};\transpo f'(t,\mathbf{x},\mathbf{z})(\tau,\pmb{\xi},\zeta)\right);\;(t,\mathbf{x},\mid\mathbf{z}\mid;\tau,\pmb{\xi},\zeta))\in WF(\Phi)\right\}
$$
therefore
\begin{equation}
WF(\Psi_0)\subset \left\{\left(t,\mathbf{x},\mathbf{z};\tau,\pmb{\xi},\zeta\frac{\mathbf{z}}{\mid\mathbf{z}\mid}\right);\;(t,\mathbf{x},\mid\mathbf{z}\mid;\tau,\pmb{\xi},\zeta))\in WF(\Phi)\right\}.
 \label{wff}
 \end{equation}
We consider also the function $g:\;\RR_t\times\RR^3_{\mathbf{x}}\times],\infty[_z\rightarrow
\RR_t\times\RR^3_{\mathbf{x}}\times \left(\RR^N_{\mathbf{z}}\setminus\{0\}\right)$ given by
$g (t,\mathbf{x},z)= \left(t,\mathbf{x},z,0_{\RR^{N-1}}\right)$. We have $\Phi=g^*\Psi_0$ and by the same theorem
$$
WF(\Phi)\subset g^*(WF(\Psi_0))=\left\{\left(t,\mathbf{x},z;\pmb{\xi},\zeta_1\right);\;\;\left(t,\mathbf{x},z,0_{\RR^{N-1}};\tau,\pmb{\xi},\pmb{\zeta}\right)\in WF(\Psi_0)\right\},
$$
hence by using (\ref{wff}) we see that $\pmb{\zeta}=\left(\zeta,0_{\RR^{N-1}}\right)$, and so
\begin{equation}
WF(\Phi)\subset\left\{\left(t,\mathbf{x},z;\tau,\pmb{\xi},\zeta\right);\;\left(t,\mathbf{x},z,0_{\RR^{N-1}};\tau,\pmb{\xi},\zeta,0_{\RR^{N-1}}\right)\in WF(\Psi_0)\right\}.
 \label{wfff}
\end{equation}
We conclude from (\ref{wff}) and (\ref{wfff}) that
\begin{equation}
\left(t,\mathbf{x},\mid z\mid;\tau,\pmb{\xi},\zeta\right)\in WF(\Phi)\Longleftrightarrow
\left(t,\mathbf{x},z,0_{\RR^{N-1}};\tau,\pmb{\xi},\zeta\frac{z}{\mid z\mid},0_{\RR^{N-1}}\right)\in WF(\Psi_0)
 \label{wfi}
\end{equation}
We now consider $\Psi(t, \mathbf{x},\mathbf{z}):=\mid\mathbf{z}\mid^{-\frac{N-1}{2}}\Psi_0(t, \mathbf{x},\mathbf{z})$. We have $WF(\Psi)=WF(\Psi_0)$. Moreover the Lemma \ref{lembo} says that $\Psi$ is solution of the free wave equation in the whole Minkowski space-time  $\RR_t\times\RR^3_{\mathbf{x}}\times\RR^N_{\mathbf{z}}$. Then the theorem of the propagation of the singularities (Theorem 8.3.3 in\cite{hormander1}) assures that
$
\left(t,\mathbf{x},z,0_{\RR^{N-1}};\tau,\pmb{\xi},\zeta\frac{z}{\mid z\mid},0_{\RR^{N-1}}\right)\in WF(\Psi)$
iff $ \forall\lambda\in\RR$, $$
\left(t+\lambda\tau,\mathbf{x}-\lambda\pmb{\xi},z-\lambda\zeta,0_{\RR^{N-1}};\tau,\pmb{\xi},\zeta\frac{z}{\mid z\mid},0_{\RR^{N-1}}\right)\in WF(\Psi).
$$
Now the result follows from (\ref{wfi}).
\fin


We end this part with some results of decay. Such properties are important to investigate the possible stability of the space-time with respect to the gravitational fluctuations. The following asymptotic behaviours are straightly deduced from the sharp estimates for the free wave equation established by P. d'Ancona, V. Georgiev and H. Kubo \cite{ancona}. It will be useful to introduce some weighted Sobolev spaces in the spirit of Y. Choquet-Bruhat  and D. Christodoulou \cite{choquet} to impose some constraints at the space infinity and at the horizon $z=0$. Given an integer $s\geq 0$ and a real $\delta$, we define the space $H^{s,\delta}_0(\RR^3_{\mathrm x}\times]0,\infty[_z)$ as the completion of $C^{\infty}_0\left(\RR^3_{\mathrm x}\times]0,\infty[_z\right)$ for the norm :
\begin{equation}
\parallel\Phi\parallel^2_{H^{s,\delta}_0}:=\sum_{\mid\alpha\mid\leq s}\sum_{k=0}^{s-\mid\alpha\mid}\sum_{b=0}^k\parallel\left(z^k+z^{b-k}\right)\left(1+\mid\mathrm x\mid^2+z^2\right)^{\frac{\delta+\mid\alpha\mid+k}{2}}\partial_{\mathrm x}^{\alpha}\partial_z^b\Phi\parallel^2_{L^2(\RR^3_{\mathrm x}\times]0,\infty[_z)}.
 \label{}
\end{equation}


\begin{Theorem}
The finite energy solution of (\ref{eq}) and (\ref{ci}) with $\mu=\frac{\nu^2-1}{4}$, $\nu\in\NN^*$, satisfies the following estimates for all $\epsilon>0$ and all ${\mathrm x}\in\RR^3$, $z>0$, provided the right members are finite :
\begin{equation}
\begin{split}
\left(1+\mid t\mid+\mid\mathrm{x}\mid+z+\mid t^2-\mid \mathrm{x}\mid^2-z^2\mid\right)^{\frac{\nu}{2}+2}z^{-\frac{\nu+1}{2}}&\mid\Phi(t,\mathrm{x},z)\mid \\
\leq&C(\epsilon)\left(\parallel\Phi_0\parallel_{H^{\left[\frac{\nu+7}{2}\right],\frac{\nu+3}{2}+\epsilon}_0}+\parallel\Phi_1\parallel_{H^{\left[\frac{\nu+5}{2}\right],\frac{\nu+5}{2}+\epsilon}_0}\right),
\end{split}
 \label{}
\end{equation}
\begin{equation}
\begin{split}
\left(1+\mid t\mid+\mid\mathrm{x}\mid+z\right)^{\frac{\nu}{2}+2} z^{-\frac{\nu+1}{2}}\mid\Phi(t,\mathrm{x},z)\mid& \\ \leq
&C(\epsilon)\left(\parallel\Phi_0\parallel_{H^{\left[\frac{\nu+7}{2}\right],-\frac{1}{2}+\epsilon}_0}+\parallel\Phi_1\parallel_{H^{\left[\frac{\nu+5}{2}\right],\frac{1}{2}+\epsilon}_0}\right),
\end{split}
 \label{}
\end{equation}
\begin{equation}
\parallel (1+\mid t\mid+\mid\mathrm x\mid+z)^{-\frac{1}{2}-\epsilon}\Phi\parallel_{L^2(\RR_t\times\RR^3_{\mathrm x}\times]0,\infty[_z)}\leq
C(\epsilon)\left(\parallel\Phi_0\parallel_{H^{1,\epsilon}_0}+\parallel\Phi_1\parallel_{H^{0,1+\epsilon}_0}\right),
 \label{}
\end{equation}
where the constant $C(\epsilon)$ is independent of $\Phi_0$ and $\Phi_1$.
 \label{strich}
 
\end{Theorem}

These are much stronger than the theorem \ref{decayLun}, since we get a uniform decay in space of $z^{-\frac{1}{2}-\sqrt{\mu+\frac{1}{4}}}\Phi(t,.)$ as $t^{-2-\sqrt{\mu+\frac{1}{4}}}$ that increases with the mass, instead of $t^{-\frac{3}{2}}$. We can expect that this rate of decay is not due to the peculiar form of the mass, and that remains true for any $\mu>-\frac{1}{4}$. These estimates are not optimal with respect to the norms that appear in the right members. The functional framework could be improved by introducing Sobolev spaces $H^{s,\delta}_0$ with non integer exponent $s$ by interpolation that could allow also  to obtain many other inequalities more precise. This work should be useful to investigate the non linear problems arising in the Anti-de Sitter cosmology.\\

{\it Proof of Theorem \ref{strich}.}
It is sufficient to treat the case $\Phi_0,\,\Phi_1\in C^{\infty}_0\left(\RR^3_{\mathrm{x}}\times]0,\infty[_z\right)$. We consider $\Psi$ defined by (\ref{psi}) with $N=\nu+2$ and $l=0$ and we put $\Psi_j(\mathrm{x},\mathrm{z}):=\mid\mathrm z\mid^{-\frac{\nu+1}{2}}\Phi_j(\mathrm{x},\mid\mathrm{z}\mid)$. We use Theorem 1.1 of \cite{ancona} with $n=\nu+5$  to get with  $d=\frac{n-1}{2}$,
\begin{equation*}
\begin{split}
\left(1+\mid t\mid+\mid\mathrm{x}\mid+\mid \mathrm{z}\mid+\mid t^2-\mid \mathrm{x}\mid^2-\mid\mathrm{z}\mid^2\mid\right)^{\frac{\nu}{2}+2}&\mid\Psi(t, \mathrm{x}, \mathrm{z})\mid\\
\leq C(\epsilon)&
\left(\parallel\Psi_0\parallel_{H^{\left[\frac{\nu+7}{2}\right],\frac{\nu+3}{2}+\epsilon}}+\parallel\Psi_1\parallel_{H^{\left[\frac{\nu+5}{2}\right],\frac{\nu+5}{2}+\epsilon}}\right),
\end{split}
\end{equation*}
and with $n=0$,
$$
\mid \Psi(t, \mathrm{x}, \mathrm{z})\mid \leq C(\epsilon)\left(1+\mid t\mid+\mid\mathrm{x}\mid+\mid\mathrm{z}\mid\right)^{-\frac{\nu}{2}-2}\left(\parallel\Psi_0\parallel_{H^{\left[\frac{\nu+7}{2}\right],-\frac{1}{2}+\epsilon}}+\parallel\Psi_1\parallel_{H^{\left[\frac{\nu+5}{2}\right],\frac{1}{2}+\epsilon}}\right),
$$
and the Theorem 1.3 of this paper with $q=2$, $\rho=-\frac{1}{2}-\epsilon$,  $\sigma=0$, $n=\nu+5$ assures that
$$
\parallel (1+\mid t\mid+\mid\mathrm x\mid+\mid\mathrm{z}\mid)^{-\frac{1}{2}-\epsilon}\Psi\parallel_{L^2(\RR_t\times\RR^3_{\mathrm x}\times\RR^N_{\mathrm z})}\leq
C(\epsilon)\left(\parallel\Psi_0\parallel_{H^{1,\epsilon}}+\parallel\Psi_1\parallel_{H^{0,1+\epsilon}}\right).
$$
Here the constant $C(\epsilon)>0$ is independent of $\Psi_0$ and $\Psi_1$ and the $H^{s,\delta}$ norms on $\RR^{3+N}$ are defined by :
\begin{equation}
\parallel\psi\parallel^2_{H^{s,\delta}}:=\sum_{\mid\alpha\mid+\mid\beta\mid\leq s}\parallel\left(1+\mid\mathrm x\mid^2+\mid\mathrm z\mid^2\right)^{\frac{\delta+\mid\alpha\mid+\mid\beta\mid}{2}}\partial_{\mathrm x}^{\alpha}\partial_{\mathrm z}^{\beta}\psi\parallel^2_{L^2(\RR^3_{\mathrm x}\times\RR^N_{\mathrm z})}.
 \label{}
\end{equation}
We can check by iteration on $\beta\in\NN^N$ that for all $\phi\in  C^{\infty}_0\left(\RR^3_{\mathrm{x}}\times]0,\infty[_z\right)$, we have
$
\partial_{\mathrm x}^{\alpha}\partial_{\mathrm z}^{\beta}\left[\mid\mathrm{z}\mid^{-\frac{N-1}{2}}\phi(\mathrm{x},\mid\mathrm{z}\mid)\right]=\sum_{finite}P_c(\mathrm{z})\mid\mathrm{z}\mid^{-\frac{N-1}{2}-a}\partial_{\mathrm x}^{\alpha}\partial_z^b\phi(\mathrm{x},\mid\mathrm{z}\mid)
$,
with
$a,b,c\in\NN$,
$P_c\in\RR[\mathrm{z}]$, $d\degree{P_c}\leq c$, $b\leq\mid\beta\mid$, $c\leq\mid\beta\mid$, $ a\leq \mid\beta\mid-b+c$.
We deduce that there exists $K>0$ such that for all $(\mathrm{x},\mathrm{z})\in\RR^{3+N}$ we have
$$
\left\vert\partial_{\mathrm x}^{\alpha}\partial_{\mathrm z}^{\beta}\left[\mid\mathrm{z}\mid^{-\frac{N-1}{2}}\phi(\mathrm{x},\mid\mathrm{z}\mid)\right]\right\vert\leq
K \mid\mathrm{z}\mid^{-\frac{N-1}{2}}\sum_{b\leq\mid\beta\mid}\left(\mid\mathrm{z}\mid^{\mid\beta\mid}+\mid\mathrm{z}\mid^{b-\mid\beta\mid}\right)\left\vert\partial_{\mathrm x}^{\alpha}\partial_z^b\phi(\mathrm{x},\mid\mathrm{z}\mid)\right\vert.
$$
We conclude that
\begin{equation}
\parallel \mid\mathrm{z}\mid^{-\frac{N-1}{2}}\phi(\mathrm{x},\mid\mathrm{z}\mid)\parallel_{H^{s,\delta}}\lesssim 
\parallel \phi\parallel_{H^{s,\delta}_0}
 \label{}
\end{equation}
and the theorem follows from the previous inequalities on $\Psi$.

\fin
We end these results on the asymptotic behaviours with some global estimates of Strichartz type. We present just the cases for which the energy allows to control the $L^p$-norms.

\begin{Theorem}
 \label{stiqq}
The finite energy solution of (\ref{eq}) and (\ref{ci}) with $\mu=\frac{\nu^2-1}{4}$, $\nu\in\NN^*$, satisfies the following estimate

\begin{equation}
 \parallel z^{(\nu+1)\left(\frac{1}{r}-\frac{1}{2}\right)}\Phi\parallel_{L^q\left(\RR_t;L^r(\RR^3_{\mathrm x}\times]0,\infty[_z)\right)}\lesssim\left(
\parallel\Phi_0\parallel_{BL^1_0(\RR^3_{\mathrm x}\times]0,\infty[_z)}+\parallel\Phi_1\parallel_{L^2(\RR^3_{\mathrm x}\times]0,\infty[_z)}\right)
 \label{striqun}
\end{equation} 
when
\begin{equation}
2\leq q,\;\;\frac{1}{q}+\frac{\nu+5}{r}=\frac{\nu+3}{2},\;\;\frac{1}{q}+\frac{\nu+4}{2r}\leq\frac{\nu+4}{4}.
 \label{stricun}
\end{equation}

Moreover, if $\nu=2k$, $k\in\NN^*$, then
\begin{equation}
 \parallel z^{\left(\frac{1}{r}-\frac{1}{2}\right)}\Phi\parallel_{L^q\left(\RR_t;L^r(\RR^3_{\mathrm x}\times]0,\infty[_z)\right)}\lesssim\left(
\parallel\Phi_0\parallel_{BL^1_0(\RR^3_{\mathrm x}\times]0,\infty[_z)}+\parallel\Phi_1\parallel_{L^2(\RR^3_{\mathrm x}\times]0,\infty[_z)}\right)
 \label{striqunn}
\end{equation} 
when
\begin{equation}
2\leq q,\;\;\frac{1}{q}+\frac{5}{r}=\frac{3}{2},\;\;\frac{1}{q}+\frac{2}{r}\leq 1,
 \label{stricunn}
\end{equation}
 and if $\nu=2k+1$, $k\in\NN$, then
\begin{equation}
 \parallel z^{\left(\frac{2}{r}-1\right)}\Phi\parallel_{L^q\left(\RR_t;L^r(\RR^3_{\mathrm x}\times]0,\infty[_z)\right)}\lesssim\left(
\parallel\Phi_0\parallel_{BL^1_0(\RR^3_{\mathrm x}\times]0,\infty[_z)}+\parallel\Phi_1\parallel_{L^2(\RR^3_{\mathrm x}\times]0,\infty[_z)}\right)
 \label{striqunnn}
\end{equation} 
when
\begin{equation}
2\leq q,\;\;\frac{1}{q}+\frac{6}{r}=2,\;\;\frac{1}{q}+\frac{5}{2r}\leq \frac{5}{4}.
 \label{stricunnn}
\end{equation}
\end{Theorem}

In the case of the gravitational fluctuations, we have $\nu=4$ and we can control  the weighted global $L^q$ norms by the energy :
\begin{equation*}
\parallel z^{-\frac{3}{4}}\Phi\parallel_{L^{\frac{20}{7}}\left(\RR_t\times\RR^3_{\mathrm x}\times]0,\infty[_z\right)}
+
\parallel z^{-\frac{1}{4}}\Phi\parallel_{L^4\left(\RR_t\times\RR^3_{\mathrm x}\times]0,\infty[_z\right)}\lesssim\left(
\parallel\Phi_0\parallel_{BL^1_0(\RR^3_{\mathrm x}\times]0,\infty[_z)}+\parallel\Phi_1\parallel_{L^2(\RR^3_{\mathrm x}\times]0,\infty[_z)}\right).
 \label{}
\end{equation*}
For the electromagnetic fluctuations we have $\nu=2$ and
\begin{equation*}
\parallel z^{-\frac{9}{16}}\Phi\parallel_{L^{\frac{16}{5}}\left(\RR_t\times\RR^3_{\mathrm x}\times]0,\infty[_z\right)}
+
\parallel z^{-\frac{1}{4}}\Phi\parallel_{L^4\left(\RR_t\times\RR^3_{\mathrm x}\times]0,\infty[_z\right)}\lesssim\left(
\parallel\Phi_0\parallel_{BL^1_0(\RR^3_{\mathrm x}\times]0,\infty[_z)}+\parallel\Phi_1\parallel_{L^2(\RR^3_{\mathrm x}\times]0,\infty[_z)}\right).
 \label{}
\end{equation*}


{\it Proof of Theorem \ref{stiqq}.}
We recall that the famous Strichartz estimates sharpened in \cite{keel}, state that the finite energy solutions $\Psi$ of the wave equation on the Minkowski space-time $\RR^{1+n}$ belongs to $L^q\left(\RR_t;L^r(\RR^n)\right)$ iff
\begin{equation}
\frac{1}{q}+\frac{n}{r}=\frac{n-2}{2},\;\;\frac{1}{q}+\frac{n-1}{2r}\leq\frac{n-1}{4},
 \label{constkeel}
\end{equation}
and this norm is controled by the energy :
$$
\parallel\Psi\parallel_{L^q\left(\RR_t;L^r(\RR^n)\right)}\lesssim\left(\parallel\Psi(0,.)\parallel_{BL^1(\RR^n)}+\parallel\partial_t\Psi(0,.)\parallel_{L^2(\RR^n)}\right).
$$
When $\Psi$ is given by (\ref{psi}), where $N$ and $\mu$ are related by (\ref{mukg}), we have $n=N+3$ and
$$
\parallel\Psi\parallel_{L^q\left(\RR_t;L^r(\RR^{N+3})\right)}=\parallel Y_{l,m}\parallel_{L^r(S^{N-1})} \parallel z^{(N-1)\left(\frac{1}{r}-\frac{1}{2}\right)}\Phi\parallel_{L^q\left(\RR_t;L^r(\RR^3_{\mathrm x}\times]0,\infty[_z)\right)},
$$
$$
\parallel\Psi(0,.)\parallel_{BL^1(\RR^{N+3})}=\parallel\Phi_0\parallel_{BL^1_0(\RR^3_{\mathrm x}\times]0,\infty[_z)},\;\;
\parallel\partial_t\Psi(0,.)\parallel_{L^2(\RR^{N+3})}=\parallel\Phi_1\parallel_{L^2(\RR^3_{\mathrm x}\times]0,\infty[_z)}.
$$
We deduce that
\begin{equation}
 \parallel z^{(N-1)\left(\frac{1}{r}-\frac{1}{2}\right)}\Phi\parallel_{L^q\left(\RR_t;L^r(\RR^3_{\mathrm x}\times]0,\infty[_z)\right)}\lesssim\left(
\parallel\Phi_0\parallel_{BL^1_0(\RR^3_{\mathrm x}\times]0,\infty[_z)}+\parallel\Phi_1\parallel_{L^2(\RR^3_{\mathrm x}\times]0,\infty[_z)}\right).
 \label{ineqst}
\end{equation}
(\ref{mukg}) allows to choose $N$ and $l$ such that
$$
\nu=N+2l-2,\;\;N\geq 3\;\;and\;\;l\geq 0,\;\;or\;\;N=2\;\;and\;\;l\geq 1.
$$

First we choose $l=0$, and $N=\nu+2$. Then (\ref{striqun}) and (\ref{stricun}) follow from (\ref{ineqst}) and (\ref{constkeel}) with $n=\nu+5$. Now when $\nu=2k$, $k\in\NN^*$, we take $l=k$, $N=2$ hence (\ref{striqunn}) and (\ref{stricunn}) follow from (\ref{ineqst}) and (\ref{constkeel}) with $n=5$.  Finally when $\nu=2k+1$, $k\in\NN$, we take $l=k$, $N=3$ hence (\ref{striqunnn}) and (\ref{stricunnn}) follow from (\ref{ineqst}) and (\ref{constkeel}) with $n=6$.

\fin

\section{Normalizable solutions in Brane Cosmology}
 
In brane cosmology, the Minkowski space-time $\RR_t\times\RR^3_{\mathrm x}$ is considered as a brane that is the boundary $\RR_t\times\RR^3_{\mathrm x}\times\{z=1\}$ of a part $\mathcal B$ of $AdS^5$ called the bulk. The choice of the bulk depends on the tension of this brane (see \cite{man}).  The $RS2$ Randall-Sundrum model that we have investigated in \cite{RS}, deals with the Minkowski brane with a positive tension associated to the bulk $\mathcal B=\RR_t\times\RR^3_{\mathrm x}\times]1,\infty[_z$. In this part we consider the case of the Minkowski brane with a negative tension. In this case $\mathcal B=\RR_t\times\RR^3_{\mathrm x}\times]0,1[_z$ and we have to study the Klein-Gordon equation
\begin{equation}
\left(\partial_t^2-\Delta_{\mathbf{x}}-\partial_z^2+\frac{\mu}{z^2}\right)\Phi=0,\;\;(t,{\mathbf x},z)\in\RR\times\RR^3\times]0,1[.
  \label{eqb}
\end{equation}
The boundary condition on the brane is associated to the $Z_2$ symmetry (see \cite{man} and the Appendix) that yields to the Neumann condition on the fields. With the change of unknown, we finally impose the Robin condition :
\begin{equation}
\partial_z\Phi(t,\mathbf{x},1)+\frac{3}{2}\Phi(t,\mathbf{x},1)=0,\;\;t\in\RR,\;\;\mathbf{x}\in\RR^3.
  \label{bcb}
\end{equation}
Associated to these constraints, there exists a  formally
conserved energy
\begin{equation}
E_1(\Phi,t):=\int_{\RR^3}\int_0^1\mid\nabla_{t,{\mathbf x},z}\Phi(t,{\mathbf
  x},z)\mid^2+\frac{\mu}{z^2}\mid\Phi(t,{\mathbf x},z)\mid^2d{\mathbf x}dz+\frac{3}{2}\int_{\RR^3}\Phi(t,\mathbf{x},1)d\mathbf{x}.
  \label{enerb}
\end{equation}
To solve the mixed problem in a suitable functional
framework, since $H^1\left(\RR^3_{\mathbf{x}}\times]0,1[_z\right)$ is a subspace of $C^0\left([0,1]_z;H^{\frac{1}{2}}\left(\RR^3_{\mathrm x}\right)\right)$, we can introduce the space
\begin{equation}
W^1:=\left\{\phi\in H^1\left(\RR^3_{\mathbf{x}}\times]0,1[_z\right);\;\;\phi(\mathbf{x},0)=0\right\},
  \label{}
\end{equation}
and we put
\begin{equation}
\|\phi\|_{W^1}^2:=\int_{\RR^3}\int_0^1\mid\nabla_{t,{\mathbf x},z}\phi({\mathbf
  x},z)\mid^2+\frac{\mu}{z^2}\mid\phi(t,{\mathbf x},z)\mid^2d{\mathbf x}dz+\frac{3}{2}\int_{\RR^3}\phi(\mathbf{x},1)d\mathbf{x}.
  \label{}
\end{equation}
Thanks to the Hardy inequality (\ref{hardy}), and the continuity of the trace on $z=1$,  $\|.\|_{W^1}$ is a norm
on $W^1$ when $\mu>-\frac{1}{4}$, that is equivalent to the usual
$H^1$-norm. Since $W^1$ is a closed subspace of $
H^1\left(\RR^3_{\mathbf{x}}\times]0,1[_z\right)$, we can see that
$W^1$ endowed with the norm $\|.\|_{W^1}$ is a Hilbert space. We
need to use also the space
\begin{equation}
W^2:=\left\{\phi\in W^1;\;\;\Delta_{\mathbf{x},z}\phi-\frac{\mu}{z^2}\phi\in L^2\left(\RR^3_{\mathbf{x}}\times]0,1[_z\right),\;\;\partial_z\phi(\mathbf{x},1)+\frac{3}{2}\phi(\mathbf{x},1)=0\right\}.
  \label{wdeux}
\end{equation}
This definition makes sense since if $\phi\in W^2$, then $\phi\in
H^1\left(\RR^3_{\mathbf{x}}\times]0,1[_z\right)$ and
$\Delta_{\mathbf{x},z}\phi\in
L^2\left(\RR^3_{\mathbf{x}}\times]a,1[_z\right)$ for all
$a\in]0,1[$. As a consequence, $\phi$ satisfies
$\partial_z\phi\in
C^0\left([a,1]_z;H^{-\frac{1}{2}}\left(\RR^3_{\mathbf{x}}\right)\right)$
hence the boundary condition on the brane $z=1$ is well defined.
Moreover, $W^2$ is a Hilbert space for the norm
\begin{equation}
\|\phi\|^2_{W^2}:=\|\phi\|_{W^1}^2+\|\Delta_{\mathbf{x},z}\phi-\frac{\mu}{z^2}\phi\|_{L^2}^2.
  \label{}
\end{equation}
When $\phi\in W^1$ but $\Delta_{\mathrm{x},z}\phi\notin L^2$, the trace $\partial_z\phi(\mathrm{x},1)$ does not exist. In order to the boundary condition (\ref{bcb}) makes sense, we introduce the space of the $W^2$-valued distributions on
$\RR_t$, $\mathcal{D}'(\RR_t;W^2)$, that is the set of the linear
continuous maps from $C^{\infty}_0(\RR_t)$ to $W^2$. We say that the boundary condition (\ref{bcb}) is satisfied by $\Phi$ when $\Phi\in \mathcal{D}'(\RR_t;W^2)$.

\begin{Theorem}
Given $\Phi_0\in W^1$, $\Phi_1\in L^2\left(\RR^3_{\mathbf{x}}\times]0,1[_z\right)$, there exists a
unique solution $\Phi$ of (\ref{eqb}) satisfying
\begin{equation}
\Phi\in \mathcal{D}'(\RR_t;W^2)\cap C^0\left(\RR_t;W^1\right)\cap
C^1\left(\RR_t;L^2\left(\RR^3_{\mathbf{x}}\times]0,1[_z\right)\right),
  \label{regb}
\end{equation}
\begin{equation}
\Phi(0,\mathbf{x},z)=\Phi_0(\mathbf{x},z),\;\;\partial_z\Phi(0,\mathbf{x},z)=\Phi_1(\mathbf{x},z),\;\;(\mathbf{x},z)\in\RR^3\times]0,1[.
  \label{cib}
\end{equation}
Moreover the energy of $\Phi$ is conserved :
\begin{equation}
\forall t\in\RR,\;\;E_1(\Phi,t)=E_1(\Phi,0).
  \label{conservb}
\end{equation}
When $\Phi_0\in W^2$ and $\Phi_1\in W^1$, then 
\begin{equation}
\Phi\in
C^0\left(\RR_t;W^2\right)\cap C^1\left(\RR_t;W^1\right)\cap C^2\left(\RR_t;L^2\left(\RR^3_{\mathbf{x}}\times]0,1[_z\right)\right).
  \label{regbb}
\end{equation}

  \label{PBMIXT}
\end{Theorem}

{\it Proof of Theorem \ref{PBMIXT}.} Since $\mu$ is a real number, it is sufficient to
consider just the real valued solutions, hence, in this proof, we
suppose that all the functions in $L^2$ are real valued and all the
spaces are real Hilbert spaces. We introduce the operator on $W^1\times L^2\left(\RR^3_{\mathbf
    x}\times]0,1[_z\right)$, given by 
$$
A:=\left(
\begin{array}{cc}
0&1\\
\Delta_{\mathbf{x}}+\partial_z^2-\frac{\mu}{z^2}&0
\end{array}
\right),\;\;Dom(A)=W^2\times W^1.
$$
We
prove that $A$ and $-A$ are maximal monotone operators. Given $u\in
W^2$, $v\in W^1$, we get with the Green formula
\begin{equation*}
\begin{split}
\left<A\left(
\begin{array}{c}
u\\v
\end{array}
\right),
\left(
\begin{array}{c}
u\\v
\end{array}
\right)
\right>_{W^1\times L^2}&=\left<\partial_zu(.,1),v(.,1)\right>_{H^{-\frac{1}{2}}(\RR^3),H^{\frac{1}{2}}(\RR^3)}+\frac{3}{2}\int_{\RR^3}v(\mathbf{x},1)u(\mathbf{x},1)d\mathbf{x}
\\
&=\frac{3}{2}\int_{\RR^3}v(\mathbf{x},1)u(\mathbf{x},1)-u(\mathbf{x},1)v(\mathbf{x},1)d\mathbf{x}\\
&=0.
\end{split}
\end{equation*}
On the other hand, given $f\in W^1$, $g\in L^2$, $(u_{\pm},v_{\pm})\in
W^2\times W^1$ is solution of 
$$
(A\pm Id)\left(
\begin{array}{c}
u_{\pm}\\
v_{\pm}
\end{array}
\right)
=\left(
\begin{array}{c}
f\\
g
\end{array}
\right)
$$
iff 
$$
u_{\pm}\in W^2,\;\; v=f\mp
u_{\pm},\;\;\Delta_{\mathbf{x},z}u_{\pm}-\left(\frac{\mu}{z^2}+1\right)u_{\pm}=g\mp
f.
$$
This partial differential equation is easily solved by a variational
method. We introduce the bilinear form
$$
a(u,u'):=\int_{\RR^3}\int_0^1\nabla_{\mathbf{x},z}u.\nabla_{\mathbf{x},z}u'+\left(\frac{\mu}{z^2}+1\right)uu'd\mathbf{x}dz+\frac{3}{2}\int_{\RR^3}u(\mathbf{x},1)u'(\mathbf{x},1)d\mathbf{x}.
$$
Since $a(.,.)$ is continuous and coercive on $W^1$, the Lax-Milgram
theorem implies there exists $u_{\pm}\in W^1$ such that
$$
\forall u'\in W^1,\;\;a(u_{\pm},u')=\int_{\RR^3}(g\mp f)u'(\mathbf{x},1)d\mathbf{x}.
$$
Taking $u'$ any test function, we deduce that $\Delta_{\mathbf{x},z}u_{\pm}-\left(\frac{\mu}{z^2}+1\right)u_{\pm}=g\mp
f$, and so $\Delta_{\mathbf{x},z}u_{\pm}-\frac{\mu}{z^2}u_{\pm}\in
L^2$. Then the Green formula gives for all  $u'\in W^1$
$$
a(u_{\pm},u')=\int_{\RR^3}\int_0^1\left(\Delta_{\mathbf{x},z}u_{\pm}-\left(\frac{\mu}{z^2}+1\right)u_{\pm}\right)u'd\mathbf{x}dz+\int_{\RR^3}\left(\partial_zu_{\pm}(\mathbf{x},1)+\frac{3}{2}u_{\pm}(\mathbf{x},1)\right)u'(\mathbf{x},1)d\mathbf{x}.
$$
We deduce that
$\partial_zu_{\pm}(\mathbf{x},1)+\frac{3}{2}u_{\pm}(\mathbf{x},1)=0$
and $u_{\pm}\in W^2$. We conclude that $\pm A$ is a densely defined
maximal monotone operator on $W^1\times L^2$,
hence the Hille-Yoshida-Phillips theorem assures the existence of a
unitary group $e^{tA}$ on $W^1\times L^2$, so that $\Phi$ defined by
$$
\left(
\begin{array}{c}
\Phi\\
\partial_t\Phi
\end{array}
\right):=e^{tA}\left(
\begin{array}{c}
\Phi_0\\
\Phi_1
\end{array}
\right)
$$
satisfies (\ref{eqb}), (\ref{regbb}), (\ref{cib}) and (\ref{conservb})
when $(\Phi_0,\Phi_1)\in W^2\times W^1$. When $(\Phi_0,\Phi_1)\in
W^1\times L^2$, we have to show that $\Phi\in
\mathcal{D}'(\RR_t;W^2)$. We pick a sequence  $(\Phi_0^n,\Phi_1^n)\in
W^2\times W^1$ that tends to  $(\Phi_0,\Phi_1)$ in
$W^1\times L^2$ as $n\rightarrow\infty$. Then $
\left(
\begin{array}{c}
\Phi^n\\
\partial_t\Phi^n
\end{array}
\right):=e^{tA}\left(
\begin{array}{c}
\Phi_0^n\\
\Phi_1^n
\end{array}
\right)
$ tends to $(\Phi,\partial_t\Phi)$ in $C^0\left(\RR_t;W^1\right)\cap
C^1\left(\RR_t;L^2\left(\RR^3_{\mathbf{x}}\times]0,1[_z\right)\right)$.
Thus given $\theta\in C^{\infty}_0(\RR_t)$, we have
$$
\int\theta(t)\Phi^n(t)dt\rightarrow\int\theta(t)\Phi(t)dt\;\;in\;\;W^1.
$$
Moreover we have
$$
\left(\Delta_{\mathbf{x},z}-\frac{\mu}{z^2}\right)\int\theta(t)\Phi^n(t)dt=
\int\theta''(t)\Phi^n(t)dt\rightarrow\int\theta''(t)\Phi(t)dt\;\;in\;\;L^2.
$$
We conclude that $\int\theta(t)\Phi^n(t)dt$ is a Cauchy sequence in
$W^2$ and $\theta\mapsto \int\theta(t)\Phi(t)dt$ is a $W^2$-valued
distribution on $\RR_t$.

To achieve the proof of the theorem and to establish the uniqueness,  we show that any $\Phi$ solution
of (\ref{eqb}) and (\ref{regb}), satisfies also (\ref{conservb}). We
take $\theta\in C^{\infty}_0(\RR)$, $0\leq\theta$,
$\int\theta(t)dt=1$, and we put
$\Phi^n(t):=n\int\theta(ns)\Phi(t+s)ds$. $\Phi^n$ is a solution of
(\ref{eqb}) and belongs to $C^{\infty}(\RR_t;W^2)$. Then we can
multiply the PDE by $\partial_t\Phi^n$ and integrate on $\RR^3_{\mathbf
  x}\times]0,1[_z$ to obtain that $E_1(\Phi^n,t)=E_1(\Phi^n,0)$. Since
$\Phi^n$ tends to $\Phi$ in  $C^0\left(\RR_t;W^1\right)\cap
C^1\left(\RR_t;L^2\left(\RR^3_{\mathbf{x}}\times]0,1[_z\right)\right)$
as $n\rightarrow\infty$,
we conclude that  $E_1(\Phi,t)=E_1(\Phi,0)$.

\fin

Following C.H. Wilcox \cite{wilcox}, the fields satisfying
(\ref{regb}) (respect. (\ref{regbb})), are called {\it finite energy
  solutions} (respect. {\it strict finite energy solutions}). They
belong to a larger class of solutions, the so-called {\it normalizable
  solutions} that are the fields that are  square integrable in space
at each time. When the coefficients are smooth up to the boundary,
these weak solutions have been studied by Vishik and Ladyzhenskaya
\cite{vishik} and in an abstract setting that allows a time dependence
of the coefficients
by J-L Lions (chapter 3 of \cite{lions}). To define the space of the initial velocity, we introduce the space
$W^{-1}$ defined as the dual space of $W^1$, endowed with its
canonical norm. We warn that the elements
of this space are not distributions since $C^{\infty}_0(\RR^3_{\mathbf
  x}\times ]0,1[_z)$ is not dense in $W^1$ ; nevertheless since $W^1$
is dense in $L^2(\RR^3_{\mathbf
  x}\times ]0,1[_z)$, this space can be identified with a subspace of $W^{-1}$. 

\begin{Theorem}
Given $\Phi_0\in L^2\left(\RR^3_{\mathbf{x}}\times]0,1[_z\right)$, $\Phi_1\in W^{-1}$, there exists a
unique solution $\Phi$ of (\ref{eqb}), (\ref{cib}) satisfying
\begin{equation}
\Phi\in \mathcal{D}'(\RR_t;W^2)\cap
C^0\left(\RR_t;L^2\left(\RR^3_{\mathbf{x}}\times]0,1[_z\right) \right)\cap C^1\left(\RR_t;W^{-1}\right).
  \label{regbw}
\end{equation}
Furthermore $\Phi$ satisfies for all $t\in\RR$ :
\begin{equation}
\|\partial_t\Phi(t)\|_{W^{-1}}^2+\|\Phi(t)\|_{L^2}^2=\|\Phi_1\|_{W^{-1}}^2+\|\Phi_0\|_{L^2}^2.
  \label{conservw}
\end{equation}
\label{normso}
\end{Theorem}

{\it Proof of Theorem \ref{normso}.} We consider the densely defined operator $L$ on
$L^2\left(\RR^3_{\mathbf{x}}\times]0,1[_z\right)$ defined by
\begin{equation}
L:=-\Delta_{\mathbf{x}}-\partial_z^2+\frac{\mu}{z^2},\;\;Dom(L)=W^2.
  \label{L}
\end{equation}

$L$ is symmetric and thanks to the Hardy estimate we have
\begin{equation}
\mu+\frac{1}{4}\leq L.
  \label{}
\end{equation}
Moreover we have shown in the proof of the previous theorem that $L+1$
is an isomorphism from $W^2$ onto $L^2$. Therefore $L$ is selfadjoint
on $L^2$, $Dom(L^{\frac{1}{2}})=W^1$ and $L^{-\frac{1}{2}}$ can be
uniquely extended in an isometry, again denoted $L^{-\frac{1}{2}}$,
from $W^{-1}$ onto $L^2$. Then $\Phi$ given by
\begin{equation}
\Phi(t):=\cos\left( tL^{\frac{1}{2}}\right)\Phi_0+\sin\left( tL^{\frac{1}{2}}\right)L^{-\frac{1}{2}}\Phi_1
  \label{}
\end{equation}
belongs to
$C^0\left(\RR_t;L^2\left(\RR^3_{\mathbf{x}}\times]0,1[_z\right)
\right)\cap C^1\left(\RR_t;W^{-1}\right)$ and satisfies  (\ref{eqb}),
(\ref{cib}) and (\ref{conservw}). To prove that this function belongs
to $\mathcal{D}'(\RR_t;W^2)$, we proceed like above.  We pick a sequence  $(\Phi_0^n,\Phi_1^n)\in
W^2\times W^1$ that tends to  $(\Phi_0,\Phi_1)$ in
$L^2\times W^{-1}$ as $n\rightarrow\infty$. Then $
\Phi^n(t):=\cos\left( tL^{\frac{1}{2}}\right)\Phi_0^n+\sin\left( tL^{\frac{1}{2}}\right)L^{-\frac{1}{2}}\Phi_1^n
$ tends to $\Phi$ in $C^0\left(\RR_t;L^2\right)$.
Thus given $\theta\in C^{\infty}_0(\RR_t)$, 
$
\int\theta(t)\Phi^n(t)dt
$
tends to
$
\int\theta(t)\Phi(t)dt
$
in $L^2$
Moreover we have
$L\int\theta(t)\Phi^n(t)dt=
\int\theta''(t)\Phi^n(t)dt\rightarrow\int\theta''(t)\Phi(t)dt\;\;in\;\;L^2.
$
We conclude that $\int\theta(t)\Phi^n(t)dt$ is a Cauchy sequence in
$W^2$ and $\theta\mapsto \int\theta(t)\Phi(t)dt$ is a $W^2$-valued
distribution on $\RR_t$.
To  establish the uniqueness,  we show that any $\Phi$ solution
of (\ref{eqb}) and (\ref{regbw}), satisfies also (\ref{conservw}). We
take $\theta\in C^{\infty}_0(\RR)$, $0\leq\theta$,
$\int\theta(t)dt=1$, and we put
$\Phi^n(t):=n\int\theta(ns)\Phi(t+s)ds$. $\Phi^n$ is a solution of
(\ref{eqb}) and belongs to $C^{\infty}(\RR_t;W^2)$. Then we can
multiply the PDE by $L^{-1}\partial_t\Phi^n$ and integrate on $\RR^3_{\mathbf
  x}\times]0,1[_z$ to obtain that $
\|\partial_t\Phi^n(t)\|_{W^{-1}}^2+\|\Phi^n(t)\|_{L^2}^2=\|\Phi_1^n\|_{W^{-1}}^2+\|\Phi_0^n\|_{L^2}^2$. Since
$\Phi^n$ tends to $\Phi$ in  $C^0\left(\RR_t;L^2\right)\cap
C^1\left(\RR_t;W^{-1}\right)$
as $n\rightarrow\infty$,
we get (\ref{conservw}).

\fin

Now we express the fields as an expansion of massive Klein-Gordon fields propagating in the Minkowski space-time, the so called {\it Kaluza-Klein tower}. Unlike the case of the positive-tension brane investigated in \cite{RS}, for which there exists a continuum of modes, the mass of the modes for the negative-tension brane is quantized and there is no massless gravito. This fact is due to the boundedness of the depth since $z\in]0,1[$ instead of $z\in]1,\infty[$.

\begin{Theorem}
There exists a sequence $(\lambda_n)_{n\in\NN}\subset]0,\infty[$ with
$\lim_{n\rightarrow\infty}\lambda_n=\infty$, and a Hilbert basis of $L^2(0,1)$,  $\left(u_n\right)_{n\in\NN}\subset
H^1(]0,1[)\cap C^{\infty}(]0,1])$ with $u_n(0)=0$,
$u'_n(1)+\frac{3}{2}u_n(1)=0$, such that for any $\Phi_0\in
L^2(\RR^3_{\mathbf x}\times]0,1[_z)$, $\Phi_1\in W^{-1}$, the
normalizable solution $\Phi$ of (\ref{eqb}), (\ref{cib}), (\ref{regbw})
can be written as 
\begin{equation}
\Phi(t,\mathbf{x},z)=\sum_0^{\infty}\phi_n(t, \mathbf{x})u_n(z)
  \label{exp}
\end{equation}
where $\phi_n\in C^0(\RR_t;L^2(\RR^3_{\mathbf x}))\cap
C^1(\RR_t;H^{-1}(\RR^3_{\mathbf x}))$ is solution of the Klein-Gordon
equation
\begin{equation}
\partial_t^2\phi_n-\Delta_{\mathbf x}\phi_n+\lambda_n^2\phi_n=0,
  \label{kglfi}
\end{equation}
and the limit (\ref{exp}) holds in 
$C^0\left(\RR_t;L^2(\RR^3_{\mathbf x}\times]0,1[_z)\right)\cap
C^1\left(\RR_t;W^{-1}\right).$
Moreover we have
\begin{equation}
\|\Phi_0\|^2_{L^2}+\|\Phi_1\|^2_{W^{-1}}=\sum_0^{\infty}\|\phi_n(t)\|_{L^2}^2+\|(-\Delta_{\mathbf{x}}+\lambda_n^2)^{-\frac{1}{2}}\partial_t\phi_n(t)\|_{L^2}^2.
  \label{consfaible}
\end{equation}
When $\Phi$ is a finite energy solution, then  $\phi_n\in C^0(\RR_t;H^1(\RR^3_{\mathbf x}))\cap
C^1(\RR_t;L^2(\RR^3_{\mathbf x}))$, the limit (\ref{exp}) holds in 
$C^0\left(\RR_t;W^1\right)\cap
C^1\left(\RR_t;L^2(\RR^3_{\mathbf x}\times]0,1[_z)\right)$, and we
have
\begin{equation}
\|\Phi_0\|^2_{W^1}+\|\Phi_1\|^2_{L^2}=\sum_0^{\infty}\|\nabla_{t,\mathbf{x}}\phi_n(t)\|_{L^2}^2+\lambda_n^2\|\phi_n(t)\|_{L^2}^2.
  \label{consfor}
\end{equation}
  \label{decompobrane}
\end{Theorem}

The sequences $\lambda_n$ and $u_n$ are explicitly given by the formulas (\ref{lambdan}) and (\ref{unn}) below. In particular, in the case of the gravitational fluctuations for which $\mu=\frac{15}{4}$, $\lambda_n$ is the set of the strictly positive zeros of the Bessel function $J_1(x)$ and $u_n(z)=\sqrt{2z}\frac{J_2(\lambda_nz)}{J_2(\lambda_n)}$.

First we develop the spectral analysis of the
one-dimensional operator
\begin{equation}
{\mathbf h}:=-\frac{d^2}{dz^2}+\frac{\mu}{z^2},\;\;Dom({\mathbf h}):=\left\{u\in
  H^1(]0,1[);\;-u''+\frac{\mu}{z^2}u\in L^2(]0,1[),\;u(0)=0,\;u'(1)+\frac{3}{2}u(1)=0\right\}.
  \label{h!}
\end{equation}

\begin{Lemma}
For all $\mu>-\frac{1}{4}$, ${\mathbf h}$ is a strictly positive selfadjoint
operator on $L^2(]0,1[)$ with a compact resolvent. Its spectrum is
formed by the sequence of simple eigenvalues $\left(\lambda_n^2\right)_{n\in\NN}$ defined by
\begin{equation}
\lambda_n>0,\;\;J_{\lambda-1}(\lambda_n)=0,\;\;\lambda:=\sqrt{\frac{1}{4}+\mu},
  \label{lambdan}
\end{equation}
associated with the $L^2$-normalized eigenfunctions
\begin{equation}
u_n(z):=C_n\sqrt{\lambda_nz}J_{\lambda}(\lambda_nz), \;\;C_n:=\sqrt{\frac{2\lambda_n}{(4+\lambda_n^2-\lambda^2)J_{\lambda}^2(\lambda_n)}}.
  \label{unn}
\end{equation}
  \label{spech}
\end{Lemma}

{\it Proof of Lemma \ref{spech}.}
We introduce the set
$$H^1_{(0)}(]0,1[):=\left\{u\in H^1(]0,1[);\;u(0)=0\right\}$$
that is a Hilbert space for the norm $\parallel u\parallel_{H^1_{(0)}}:=\parallel u'\parallel_{L^2}$ which is equivalent to the usual $H^1$ norm since when $u(0)=0$ we have
$$
\int_0^1u^2(z)dz\leq \int_0^1\frac{1}{z^2}u^2(z)dz\leq 4 \int_0^1\mid u'(z)\mid^2dz
$$
For any $u\in Dom(\mathbf{h})$, $v\in H^1_{(0)}(]0,1[)$, an integration by part gives
$$
<\mathbf{h}u,v>_{L^2}=q(u,v):=\frac{3}{2}u(1)v(1)+\int_0^1u'(z)v'(z)+\frac{\mu}{z^2}u(z)v(z)dz.
$$
Since the Hardy   inequality shows that 
$$
q(u,u)\geq \max\left(\left(\mu+\frac{1}{4}\right)\int_0^1\frac{1}{z^2}\mid u(z)\mid^2dz,\;\;\min(1,1+4\mu)\int_0^1\mid u'(z)\mid^2dz\right),
$$
we can see that $\mathbf{h}$ is a symmetric and strictly positive operator on $L^2(0,1)$.
Given $f\in L^2(0,1)$, $u_{\pm}\in Dom(\mathbf{h})$ is solution of $\mathbf{h}u_{\pm}\pm iu_{\pm}$ if and only if for all $v\in H^1_{(0)}(]0,1[)$ we have
$a_{\pm}(u_{\pm},v):=q (u_{\pm},v)\pm i<u_{\pm},v>_{L^2}=<f,v>_{L^2}$. The previous inequalities imply that $a_{\pm}$ is continuous and coercive on $H^1_{(0)}$ hence 
the Lax-Milgram lemma assures that $u_{\pm}$ exists, and we conclude that $\mathbf{h}$ is a selfadjoint operator on $L^2(0,1)$, with a compact resolvent by the compact embedding $H^1(]0,1[)\subset\subset L^2(0,1)$. According to the Hilbert-Schmidt theorem, there exists a Hilbert basis of $L^2$ formed by eigenfunctions $\left(u_n\right)_{n\in\NN}$ associated to a sequence of eigenvalues $\lambda_n^2>0$. The solutions of $-u_n''+\frac{\mu}{z^2}u_n=\lambda_n^2u_n$ are given by
$$
u_n(z)=\sqrt{\lambda_n z}\left[\alpha_nJ_{\lambda}(\lambda_nz)+\beta_nY_{\lambda}(\lambda_nz)\right],\;\;\lambda:=\sqrt{\mu+\frac{1}{4}},\;\;\alpha_n,\,\beta_n\in\CC.
$$
From the asympotics , $J_{\lambda}(x)\sim x^{\lambda}$,  $Y_{\lambda}(x)\sim x^{-\lambda}$, $J_{\lambda}'(x)\sim x^{\lambda-1}$,  $Y_{\lambda}'(x)\sim x^{-\lambda-1}$ as $x\rightarrow 0^+$, we deduce that $\beta_n=0$  when $u_n\in H^1_{(0)}(]0,1[)$. To calculate the constant of normalization, we use the formula (11.4.5) of \cite{abra} with $a=2$ and $b=1$ to get
$$
\int_0^1\lambda_nzJ_{\lambda}^2(\lambda_nz)dz=\frac{1}{2\lambda_n}(4+\lambda_n^2-\lambda^2)J_{\lambda}^2(\lambda_n).
$$
Finally the condition at $z=1$ implies
$2J_{\lambda}(\lambda_n)+\lambda_nJ'_{\lambda}(\lambda_n)=0$ that is equivalent to (\ref{lambdan}). In particular $J_{\lambda}(\lambda_n)\neq 0$ and $C_n$ is well defined.
\fin 

{\it Proof of Theorem \ref{decompobrane}.}
We use the Hilbert basis $\left(u_n\right)_{n\in\NN}$ of the previous Lemma.
Given $\Phi\in C^0\left(\RR_t;L^2(\RR^3_{\mathrm{x}}\times]0,1[_z)\right)$ we have for all $t\in\RR$ and almost all $\mathrm{x}\in\RR^3$,
$$
\Phi(t,\mathrm{x},z)=\lim_{N\rightarrow\infty} \sum_{n=0}^N\phi_n(t,\mathrm{x})u_n(z)\;\;in\;\;L^2\left(]0,1[_z,dz\right),\;\;\phi_n(t,\mathrm{x}):=\int_0^1\Phi(t,\mathrm{x},z)u_n(z)dz.
$$
The Fubini theorem implies
\begin{equation}
 \parallel \Phi(t,.) \parallel^2_{L^2(\RR^3_{\mathrm x}\times]0,1[_z)}=\sum_{n=0}^{\infty}
\parallel \phi_n(t,.)\parallel^2_{L^2(\RR^3_{\mathrm x})},
 \label{nL2}
\end{equation}
and also that
$$
\parallel \phi_n(t,.)-\phi_n(s,.) \parallel_{L^2(\RR^3_{\mathrm x})}\leq \parallel \Phi(t,.)-\Phi(s,.) \parallel_{L^2(\RR^3_{\mathrm x}\times]0,1[_z)}
$$
hence $\phi_n\in C^0\left(\RR_t;L^2\left(\RR^3_{\mathrm x}\right))\right)$. Moreover
$$
\parallel \Phi(t,.)-\sum_{n=0}^N\phi_n(t,.)u_n\parallel^2_{L^2\left(\RR^3_{\mathrm x}\times]0,1[_z\right)}=
\int_{\RR^3}\parallel \Phi(t,\mathrm{x},.)-\sum_{n=0}^N\phi_n(t,\mathrm{x})u_n(.)\parallel^2_{L^2\left(]0,1[_z\right)}d\mathrm{x},
$$
and since $\parallel \Phi(t,\mathrm{x},.)-\sum_{n=0}^N\phi_n(t,\mathrm{x})u_n(z)\parallel^2_{L^2\left(]0,1[_z\right)}\leq 4 \parallel \Phi(t,\mathrm{x},.) \parallel^2_{L^2\left(]0,1[_z\right)}$, we deduce from the dominated convergence theorem that
$$
\Phi(t,\mathrm{x},z)=\lim_{N\rightarrow\infty}\sum_{n=0}^N\phi_n(t,\mathrm{x})u_n(z)\;\;in\;\;L^2\left(\RR^3_{\mathrm x}\times]0,1[_z\right)
 $$
and this limit is uniform on the compacts of $ L^2\left(\RR^3_{\mathrm{x}}\times]0,1[_z\right)$. We conclude that for any $\Phi\in C^0\left(\RR_t;L^2\left(\RR^3_{\mathrm{x}}\times]0,1[_z\right)\right)$ we have
\begin{equation}
\Phi(t,\mathrm{x},z)=\lim_{N\rightarrow\infty}\sum_{n=0}^N\phi_n(t,\mathrm{x})u_n(z)\;\;in\;\;C^0\left(\RR_t;L^2\left(\RR^3_{\mathrm x}\times]0,1[_z\right)\right).
 \label{cvL2}
\end{equation}
We can extend $L$ into an isometry $\tilde{L}$ from $W^1$ onto $W^{-1}$. Given $\varphi,\psi\in C^{\infty}_0\left(\RR^3_{\mathrm x}\right)$ and $u_n$, $u_p$, we have
$$
\left<\varphi\otimes u_n;\psi\otimes u_p\right>_{W^{-1}}=
\left<\tilde{L}^{-1}\left[\varphi\otimes u_n\right];\tilde{L}^{-1}\left[\psi\otimes u_p\right]\right>_{W^{1}}
=\left<\tilde{L}^{-1}\left[\varphi\otimes u_n\right];\psi\otimes u_p\right>_{L^2}.
$$
Since we can check that $L\left(\left[-\Delta_{\mathrm x}+\lambda_n^2\right]^{-1}\varphi\otimes u_n\right)=\varphi\otimes u_n$, we deduce that
\begin{equation*}
\begin{split}
\left<\varphi\otimes u_n;\psi\otimes u_p\right>_{W^{-1}}
&=\left<\left[-\Delta_{\mathrm x}+\lambda_n^2\right]^{-1}\varphi\otimes u_n;\psi\otimes u_p\right>_{L^2\left(\RR^3_{\mathrm x}\times]0,1[_z\right)}\\
&=\left<\left[-\Delta_{\mathrm x}+\lambda_n^2\right]^{-\frac{1}{2}}\varphi;\left[-\Delta_{\mathrm x}+\lambda_p^2\right]^{-\frac{1}{2}}\psi\right>_{L^2\left(\RR^3_{\mathrm x}\right)}\delta_{n,p},
\end{split}
\end{equation*}
where $\delta_{n,p}$ is the symbol of Kronecker. We deduce that
$$
\parallel\Phi(t,.)\parallel_{W^{-1}}^2=\sum_{n=0}^{\infty}\parallel \left[-\Delta_{\mathrm x}+\lambda_n^2\right]^{-\frac{1}{2}}\phi_n(t,.)\parallel_{L^2}^2.
$$
As a consequence of the density of $L^2$ in $W^{-1}$, given $\Phi'\in W^{-1}$, the map 
$$\phi'_n:\;\varphi\in H^1(\RR^3_{\mathrm x})\mapsto \left<\Phi';\varphi\otimes u_n\right>_{W^{-1},W^1}$$
 is a well defined distribution of $H^{-1}(\RR^3_{\mathrm x})$ and
$$
\Phi'=\lim_{N\rightarrow\infty} \sum_{n=0}^N\phi_n'\otimes u_n\;\;in\;\;W^{-1},
$$
\begin{equation}
\parallel\Phi'\parallel_{W^{-1}}^2=\sum_{n=0}^{\infty}\parallel \left[-\Delta_{\mathrm x}+\lambda_n^2\right]^{-\frac{1}{2}}\phi'_n\parallel_{L^2}^2.
 \label{nw-1}
\end{equation}
More generally, for any $\Phi'\in C^0\left(\RR_t;W^{-1}\right)$ we have
\begin{equation}
\left<\phi'_n(t);\varphi\right>_{H^{-1},H^1}:= \left<\Phi'(t);\varphi\otimes u_n\right>_{W^{-1},W^1}\in C^0\left(\RR_t\right),
 \label{}
\end{equation}
hence $\phi'_n\in C^0\left(\RR_t;H^{-1}\left(\RR^3_{\mathrm x}\right)\right)$, and 
\begin{equation}
\Phi'(t)=\lim_{N\rightarrow\infty}\sum_{n=0}^N\phi_n'(t)\otimes u_n\;\;in\;\;C^0\left(\RR_t;W^{-1}\right).
 \label{cvw-1}
\end{equation}

Similarly we have
$$
\left<\varphi\otimes u_n;\psi\otimes u_p\right>_{W^{1}}=
\left<L\left[\varphi\otimes u_n\right];\psi\otimes u_p\right>_{L^2}=\left<\nabla_{\mathrm x}\varphi;\nabla_{\mathrm x}\psi\right>_{L^2}\delta_n^p+\lambda_n^2 \left<\varphi;\psi\right>_{L^2}\delta_n^p,
$$
hence when  $\Phi\in C^0\left(\RR_t;W^1\right)$ we have
\begin{equation}
\Phi(t,\mathrm{x},z)=\lim_{N\rightarrow\infty}\sum_{n=0}^N\phi_n(t,\mathrm{x})u_n(z)\;\;in\;\;C^0\left(\RR_t;W^1\right),
 \label{cvw1}
\end{equation} 
\begin{equation}
\|\Phi(t)\|^2_{W^1}=\sum_0^{\infty}\|\nabla_{\mathbf{x}}\phi_n(t)\|_{L^2}^2+\lambda_n^2\|\phi_n(t)\|_{L^2}^2.
  \label{consw1}
\end{equation}
 \label{}
Now we consider a normalizable solution $\Phi$. To prove that $\phi_n$ defined as above is solution of the Klein-Gordon equation on the Minkowski space-time with the mass $\lambda_n^2$, we take $\chi\in C^{\infty}_0\left(\RR_t\times\RR^3_{\mathrm x}\right)$ and $\theta\in C^{\infty}_0\left(]0,1[_z\right)$, and we write :
\begin{equation}
\begin{split}
0&=\left<\left(\partial_t^2-\Delta_{\mathbf{x}}-\partial_z^2+\frac{\mu}{z^2}\right)\Phi;\chi\otimes\theta\right>_{\mathcal{D}',\mathcal{D}}\\
&=\left<\Phi;\left(\partial_t^2-\Delta_{\mathbf{x}}\right)\chi\otimes\theta+\chi\otimes\left( -\partial_z^2+\frac{\mu}{z^2}\right)\theta\right>_{L^2}\\
&=\sum_{n=0}^{\infty}\left<\phi_n;\left(\partial_t^2-\Delta_{\mathbf{x}}\right)\chi\right>_{L^2}\left<u_n;\theta\right>_{L^2}+
\left<\phi_n;\chi\right>_{L^2}\left<u_n;\left( -\partial_z^2+\frac{\mu}{z^2}\right)\theta\right>_{L^2}\\
&=\sum_{n=0}^{\infty}\left<\partial_t^2\phi_n-\Delta_{\mathbf x}\phi_n+\lambda_n^2\phi_n;\chi\right>_{\mathcal{D}',\mathcal{D}}\left<\theta;u_n\right>_{L^2},
\end{split}
 \label{}
\end{equation}
and we conclude that $\phi_n$ is solution of the Klein-Gordon equation (\ref{kglfi}). Now the expansion (\ref{exp}) is a consequence of (\ref{cvL2}) and (\ref{cvw-1}). (\ref{nL2}) and (\ref{nw-1}) imply the conservation law (\ref{consfaible}). When $\Phi$ is a finite energy solution, (\ref{cvw1}) and (\ref{cvL2}) assure the convergence of (\ref{exp}) in $C^0\left(\RR_t;W^1\right)\cap
C^1\left(\RR_t;L^2(\RR^3_{\mathbf x}\times]0,1[_z)\right)$, and the expansion of the energy (\ref{consfor})  is given by (\ref{nL2}) and (\ref{consw1}).

\fin


We end this part by  a result of uniform decay that is a very modest step toward the much more difficult study of the non-linear stability of the brane. By the way, the $L^2-L^{\infty}$ estimates are very useful to the analysis of
non-linear problems. 
We shall use  a dyadic partition of the unity  $\chi_p\in C^{\infty}_0([0,\infty[)$ satisfying
\begin{equation*}
\sum_{p=0}^{\infty}\chi_p=1,\;\;\mathrm{supp}\chi_0\subset[0,2[,\;\;1\leq
p\Rightarrow \mathrm{supp}\chi_p\subset [2^{p-1},2^{p+1}].
\end{equation*}
Given an integer $k\geq 0$, and two integer-valued functions $k'$, $k''$ defined on $\NN^3$, we introduce the functional space
$$
\Theta(k,k',k''):=\left\{\Phi\in L^2\left(\RR^3_{\mathrm x}\times]0,1[_z\right);\;\;\mid\alpha\mid\leq k
\Rightarrow
\partial_{\mathrm x}^{\alpha}\Phi\in Dom\left(L^{\max(k'(\alpha),k''(\alpha))}\right),\;\; \parallel\Phi\parallel_{\Theta(k,k',k")}<\infty\right\},
$$
where $L$ is the operator defined by (\ref{L}). On this space we define the norm :
\begin{equation*}
\begin{split}
\parallel\Phi\parallel_{\Theta(k,k',k'')}:=
&\sum_{\mid\alpha\mid\leq k}\parallel(1+\mid
  \mathrm{x}\mid)^{\frac{3}{2}}\partial_{\mathrm{x}}^{\alpha}\left(-\partial_z^2+\frac{\mu}{z^2}\right)^{k'(\alpha)}\Phi\|_{L^2(\RR^3_{\mathrm{x}}\times]0,1[_z)}\\
&+
\sum_{\mid\alpha\mid\leq k}\sum_{p=0}^{\infty}\|\chi_p(\mid\mathrm{x}\mid)(1+\mid
  \mathrm{x}\mid)^{\frac{3}{2}}\partial_{\mathrm{x}}^{\alpha}\left(-\partial_z^2+\frac{\mu}{z^2}\right)^{k''(\alpha)}\Phi\|_{L^2(\RR^3_{\mathrm{x}}\times]0,1[_z)}.
\end{split}
\end{equation*}


\begin{Theorem}
There exists $C>0$ such that for all $\Phi_0\in\Theta(3,k',k''_0)$ and $\Phi_1\in\Theta(2,k',k''_1)$, the finite energy solution $\Phi$ satisfies the following estimates :
\begin{equation}
\mid\Phi(t,\mathrm{x},z)\mid
\leq C(\mid t\mid+\mid\mathrm{x}\mid)^{-\frac{3}{2}}z^{\lambda+\frac{1}{2}}\left(\parallel\Phi_0\parallel_{\Theta(3,k',k_0'')}+\parallel\Phi_1\parallel_{\Theta(2,k',k_1'')}\right),
 \label{decayz}
\end{equation}
with
\begin{equation}
k'(\alpha)=\left[\frac{\lambda+1}{2}\right]+1,\;\;k''_j(\alpha)=\left[\frac{2\lambda+5-2\mid\alpha\mid-2j}{4}\right]+1,\;\;\lambda=\sqrt{\mu+\frac{1}{4}}
 \label{decayzk}
\end{equation}
and
\begin{equation}
\mid\Phi(t,\mathrm{x},z)\mid
\leq C(\mid t\mid+\mid\mathrm{x}\mid)^{-\frac{3}{2}}\left(\parallel\Phi_0\parallel_{\Theta(3,k',k_0'')}+\parallel\Phi_1\parallel_{\Theta(2,k',k_1'')}\right),
 \label{decayzz}
\end{equation}
with
\begin{equation}
k'(\alpha)=1,\;\;k''_j(\alpha)=\left[\frac{2-\mid\alpha\mid-j}{2}\right]+1.
 \label{decayzkk}
\end{equation}
  \label{theollde}
\end{Theorem}

{\it Proof of Theorem \ref{theollde}.}
In a first time, we assume that $\Phi_j\in C^{\infty}\left(\RR^3_{\mathrm x};L^2]0,1[_z\right)$ and  for all $\mathrm{x}\in\RR^3$, the maps $z\mapsto\partial_{\mathrm x}^{\alpha}\Phi_j(\mathrm{x},z)$ belong to $Dom\left({\mathbf h}^{\max(k'(\alpha),k''_j(\alpha))}\right)$.
We estimate $\Phi(t,\mathrm{x},z)$ by using the expansion (\ref{exp}) :
$$
\mid \Phi(t,\mathrm{x},z)\mid\leq\sum_{n=0}^{\infty}\mid \phi_n(t,\mathrm{x})\mid \mid u_n(z)\mid.
$$
To control $u_n(z)$ we need the asymptotics for the Bessel function and its zeros
(\cite{olver}
p.238, p.247) :
$$
J_{\lambda}(z)=\sqrt{\frac{2}{\pi z}}\cos\left(z-\lambda\frac{\pi}{2}-\frac{\pi}{4}\right)+O\left(z^{-\frac{3}{2}}\right),
$$
$$
\lambda_n=\pi\left(n+\frac{\lambda}{2}-\frac{3}{4}\right)+O\left(n^{-1}\right),
$$
therefore $C_n\sim\sqrt{\pi}$ and since $J_{\lambda}(x)\sim x^{\lambda}$ as $x\rightarrow 0$, we deduce that there exists $C$ independent of $n$ such that
$$
\mid u_n(z)\mid\leq Cz^{\lambda+\frac{1}{2}}\lambda_n^{\lambda+\frac{1}{2}}.
 $$
We get :
$$
\mid\Phi(t,\mathrm{x},z)\mid\leq Cz^{\lambda+\frac{1}{2}}\sum_{n=0}^{\infty}\lambda_n^{\lambda+\frac{1}{2}}\mid\phi_n(t,\mathrm{x})\mid.
$$
We estimate the Klein-Gordon fields by the Lemma 4.3 of \cite{RS} that yields to :
\begin{equation}
\begin{split}
\mid\phi_n(t,\mathrm{x})\mid &
\leq
C(1+\mid t\mid+\mid\mathrm{x}\mid)^{-\frac{3}{2}}
\sum_{\mid\alpha\mid+j\leq 3}\parallel(1+\mid
  \mathrm{y}\mid)^{\frac{3}{2}}\partial_{\mathrm{y}}^{\alpha}\phi_n^j\|_{L^2(\RR^3_{\mathrm{y}})}\\
&+
C(\mid t\mid+\mid\mathrm{x}\mid)^{-\frac{3}{2}}\sum_{\mid\alpha\mid+j\leq 3}\sum_{p=0}^{\infty}\lambda_n^{\frac{3}{2}-\mid\alpha\mid-j}\|\chi_p(\mid\mathrm{y}\mid)(1+\mid
  \mathrm{y}\mid)^{\frac{3}{2}}\partial_{\mathrm{y}}^{\alpha}\phi_n^j\|_{L^2(\RR^3_{\mathrm{y}})}
\end{split}
 \label{}
\end{equation}
where the initial data are given by
$$
\phi_n^j(\mathrm{x}):=\int_0^1\Phi_j(\mathrm{x},z)\overline{u_n(z)}dz.
$$
Since for all $\mathrm{x}\in\RR^3$ the maps $z\mapsto \partial_{\mathrm x}^{\alpha}\Phi_j(\mathrm{x},z)$ belong to $Dom\left(\mathbf{h}^k\right)$ for some $k$, we have :
$$
\phi_n^j(\mathrm{x}):=\lambda_n^{-2k}\int_0^1\left(-\partial_z+\frac{\mu}{z^2}\right)^k\Phi_j(\mathrm{x},z)\overline{u_n(z)}dz.
$$
Since $u_n$ is a Hilbert basis of $L^2(]0,1[_z)$ we can write
$$
\parallel(1+\mid
  \mathrm{y}\mid)^{\frac{3}{2}}\partial_{\mathrm{y}}^{\alpha}\phi_n^j\|_{L^2(\RR^3_{\mathrm{y}})}\leq
\lambda_n^{-2k}A_n(\alpha,j,k),
$$
$$
\sum_{n=0}^{\infty}A_n^2(\alpha,j,k)\leq
\parallel(1+\mid
  \mathrm{y}\mid)^{\frac{3}{2}}\partial_{\mathrm{y}}^{\alpha}\left(-\partial_z+\frac{\mu}{z^2}\right)^k\Phi_j\|_{L^2(\RR^3_{\mathrm{y}}\times]0,1[_z)}^2,
$$
$$
\parallel\chi_p(\mid\mathrm{y}\mid)(1+\mid
  \mathrm{y}\mid)^{\frac{3}{2}}\partial_{\mathrm{y}}^{\alpha}\phi_n^j\|_{L^2(\RR^3_{\mathrm{y}})}\leq
\lambda_n^{-2k}B_n(\alpha,j,k,p),
$$
$$
\sum_{n=0}^{\infty}B_n^2(\alpha,j,k,p)\leq
\parallel\chi_p(\mid\mathrm{y}\mid)(1+\mid
  \mathrm{y}\mid)^{\frac{3}{2}}\partial_{\mathrm{y}}^{\alpha}\left(-\partial_z+\frac{\mu}{z^2}\right)^k\Phi_j\|_{L^2(\RR^3_{\mathrm{y}}\times]0,1[_z)}^2.
$$
Now we apply the Cauchy-Schwartz inequality and we estimate $\Phi$ as follows :
\begin{equation*}
\begin{split}
\mid\Phi(t,\mathrm{x},z)\mid &
\leq
C(\mid t\mid+\mid\mathrm{x}\mid)^{-\frac{3}{2}}z^{\lambda+\frac{1}{2}}\left[
\sum_{\mid\alpha\mid+j\leq 3}\left(\sum_{n=0}^{\infty}\lambda_n^{2\lambda+1-4k}\right)^{\frac{1}{2}}\parallel(1+\mid
  \mathrm{y}\mid)^{\frac{3}{2}}\partial_{\mathrm{y}}^{\alpha}\left(-\partial_z+\frac{\mu}{z^2}\right)^k\Phi_j\|_{L^2(\RR^3_{\mathrm{y}}\times]0,1[_z)}\right.\\
&\left.+
\sum_{\mid\alpha\mid+j\leq 3}\sum_{
    p=0}^{\infty}\left(\sum_{n=0}^{\infty}\lambda_n^{2\lambda+4-2\mid\alpha\mid-2j-4k}\right)^{\frac{1}{2}}\parallel\chi_p(\mid\mathrm{y}\mid)(1+\mid
  \mathrm{y}\mid)^{\frac{3}{2}}\partial_{\mathrm{y}}^{\alpha}\left(-\partial_z+\frac{\mu}{z^2}\right)^k\Phi_j\|_{L^2(\RR^3_{\mathrm{y}}\times]0,1[_z)}\right],
\end{split}
 \label{}
\end{equation*}
where the integers $k$ depend on $\alpha$ and  $j$. To be sure the series are convergent, it is sufficient to take $k=\left[\frac{\lambda+1}{2}\right]+1$ in the first sum and $k=\left[\frac{2\lambda+5-2\mid\alpha\mid-2j}{4}\right]+1$ in the second one, so we get (\ref{decayz}) with (\ref{decayzk}).

By the same way, since $\sqrt{x}J_{\lambda}(x)\in L^{\infty}(]0,\infty[)$, we have $\sup_n\parallel u_n\parallel_{L^{\infty}(]0,1[)}<\infty$ and
$$
\mid \Phi(t,\mathrm{x},z)\mid\leq C\sum_{n=0}^{\infty}\mid \phi_n(t,\mathrm{x})\mid.
$$
As previous we obtain
\begin{equation*}
\begin{split}
\mid\Phi(t,\mathrm{x},z)\mid &
\leq
C(\mid t\mid+\mid\mathrm{x}\mid)^{-\frac{3}{2}}\left[
\sum_{\mid\alpha\mid+j\leq 3}\left(\sum_{n=0}^{\infty}\lambda_n^{-4k}\right)^{\frac{1}{2}}\parallel(1+\mid
  \mathrm{y}\mid)^{\frac{3}{2}}\partial_{\mathrm{y}}^{\alpha}\left(-\partial_z+\frac{\mu}{z^2}\right)^k\Phi_j\|_{L^2(\RR^3_{\mathrm{y}}\times]0,1[_z)}\right.\\
&\left.+
\sum_{\mid\alpha\mid+j\leq 3}\sum_{
    p=0}^{\infty}\left(\sum_{n=0}^{\infty}\lambda_n^{3-2\mid\alpha\mid-2j-4k}\right)^{\frac{1}{2}}\parallel\chi_p(\mid\mathrm{y}\mid)(1+\mid
  \mathrm{y}\mid)^{\frac{3}{2}}\partial_{\mathrm{y}}^{\alpha}\left(-\partial_z+\frac{\mu}{z^2}\right)^k\Phi_j\|_{L^2(\RR^3_{\mathrm{y}}\times]0,1[_z)}\right],
\end{split}
 \label{}
\end{equation*}
and we choose $k=1$ in the first sum and $k=\left[\frac{2-\mid\alpha\mid-j}{2}\right]+1$ in the second one, that yields to
 (\ref{decayzz}) with (\ref{decayzkk}).

To achieve the proof of the theorem, we use a procedure of regularization by taking a sequence $\theta_n\subset C^{\infty}_0\left(\RR^3_{\mathrm x}\right)$, satisfying $0\leq\theta_n$, $\mid\mathrm x\mid\geq\frac{1}{n}\Rightarrow\theta_n(\mathrm{x})=0$, $\int\theta_n(\mathrm{x})d\mathrm{x}=1$, and given $\Phi\in\Theta(k,k',k'')$, we put $\Phi_{n}(\mathrm{x},z):=\int \Phi(\mathrm{x-y},z)\theta_n(\mathrm{y})d\mathrm{y}$. It is clear that $\Phi_{n}\in C^{\infty}\left(\RR^3_{\mathrm x};L^2]0,1[_z\right)$ and  since $\partial_{\mathrm x}^{\alpha}\Phi\in Dom\left(L^{\max(k'(\alpha),k''(\alpha))}\right)$, we have $\Phi_n\in C^{\infty}\left(\RR^3_{\mathrm x};Dom\left(L^{\sup(k',k'')}\right)\right)$. That implies that for all $\mathrm{x}\in\RR^3$ and all $\alpha\in\NN^3$, the maps $z\mapsto\partial_{\mathrm x}^{\alpha}\Phi_{n}(\mathrm{x},z)$ belong to $Dom\left({\mathbf h}^{\max(k'(\alpha),k''_j(\alpha))}\right)$. To end the proof, we show that $\Phi_n$ belongs to   in $\Theta(k,k',k'')$  and tends to $\Phi$ as $n\rightarrow\infty$. This a straight consequence of the following properties that are easy to be proved. If $(1+\mid \mathrm{x}\mid)^{\frac{3}{2}}f\in L^2\left(\RR^3_{\mathrm x}\right)$ we have
$$
(1+\mid \mathrm{x}\mid)^{\frac{3}{2}}\left\vert f*\theta_n\right\vert
\leq C\left[(1+\mid \mathrm{x}\mid)^{\frac{3}{2}}\mid f\mid\right]*\theta_n\in L^2\left(\RR^3_{\mathrm x}\right),
$$
$$
\parallel (1+\mid \mathrm{x}\mid)^{\frac{3}{2}}\left(f*\theta_n-f\right)\parallel_{L^2}\leq
C \parallel \left[ (1+\mid \mathrm{x}\mid)^{\frac{3}{2}} f\right]*\theta_n-(1+\mid \mathrm{x}\mid)^{\frac{3}{2}} f\parallel_{L^2}+O\left(\frac{1}{n}\right) \underset{n\rightarrow\infty}\longrightarrow 0,
$$
$$
\parallel \chi_p(\mathrm{x})(1+\mid \mathrm{x}\mid)^{\frac{3}{2}}\left[f*\theta_n\right]\parallel_{L^2}\leq C\sum_{j=-1}^{1}
\parallel \chi_{p+j}(\mathrm{x})(1+\mid \mathrm{x}\mid)^{\frac{3}{2}}f\parallel_{L^2},
$$
and if $\sum_{p=0}^{\infty}\parallel \chi_{p}(\mathrm{x})(1+\mid \mathrm{x}\mid)^{\frac{3}{2}}f\parallel_{L^2}<\infty$, then
$\sum_{p=0}^{\infty}\parallel \chi_{p}(\mathrm{x})(1+\mid \mathrm{x}\mid)^{\frac{3}{2}}\left[f*\theta_n\right]\parallel_{L^2}<\infty$ and
$$
\sum_{p=0}^{\infty}\parallel \chi_{p}(\mathrm{x})(1+\mid \mathrm{x}\mid)^{\frac{3}{2}}\left(f*\theta_n-f\right)\parallel_{L^2}
 \underset{n\rightarrow\infty}\longrightarrow 0.
$$
\fin

\section{Appendix}
The Anti-de-Sitter space in 4 + 1 dimensions of constant curvature $-k<0$ can be represented as the quadric
$$
X_0^2 + X_5^2 -\sum_{i=1}^4X_i^2 = \frac{1}{k^2},
$$
embedded in the 4+2 dimensional flat space $\RR_{\mathrm X}^6$ with metric
$$
ds^2 = dX_0^2 + dX_5^2 -\sum_{i=1}^4X_i^2.
$$
This metric is a maximally symmetric solution without singularity, of the Einstein equations in the vacuum, with the negative cosmological constant $-k<0$, moreover this solution is unique, for the given topology, according to the Birkhoff theorem.

We can describe this manifold with a unique chart by using the so-called global coordinates $\tau,\rho,\Omega$ defined by the relations 
$$
X_0=\frac{1}{k}\sec\rho\cos\tau,\;\;X_5=\frac{1}{k}\sec\rho\sin\tau,\;\;X_i=\frac{1}{k}\tan\rho\,\Omega_i,\;\;i=1,..,4,\;\;\sum_{i=1}^4\Omega_i^2=1.
$$
Then the metric has the form 
$$
ds^2=\frac{1}{(k\cos\rho)^2}\left[d\tau^2-d\rho^2-(\sin\rho)^2 d\Omega_{S^3}^2\right],\;\;\tau\in[-\pi,\pi],\;\;\rho\in[0,\frac{\pi}{2}[,\;\;\Omega\in S^3.
$$
Here, the points at $\tau=\pi$ and $\tau=-\pi$ have to be identified therefore this universe is totally vicious as regards the causality, and we prefer to work with $\tau\in\RR$ that corresponds to the universal covering $CAdS^5$. Whatever choice for $\tau$,  there is a time-like boundary at $\rho=\frac{\pi}{2}$, the ``horizon''.

In the coordinates, the d'Alembertian
$$
\square_g:=\frac{1}{\sqrt{\mid g\mid}}\frac{\partial}{\partial
  x^{\mu}}\left(\sqrt{\mid g\mid}g^{\mu\nu}\frac{\partial}{\partial
  x^{\nu}}\right)
$$
is expressed as :
$$
\square_g=(k\cos\rho)^2\left[\partial_{\tau}^2-\partial_{\rho}^2-\frac{3}{\cos\rho\sin\rho}\partial_{\rho}-\frac{1}{\sin^2\rho}\Delta_{S^3}\right].
$$
To cancel the first derivative with respect to $\rho$ we make a change of unknown and the Klein-Gordon equation becomes :
$$
\left(\square_g+\lambda k^2\right)u=(k\cos\rho)^2(\tan\rho)^{-\frac{3}{2}}\left[\partial_{\tau}^2-\partial_{\rho}^2+\frac{3}{4\sin^2\rho}+\frac{\frac{15}{4}+\lambda}{\cos^2\rho}-\frac{1}{\sin^2\rho}\Delta_{S^3}\right]\left((\tan\rho)^{\frac{3}{2}} u\right).
$$
Following \cite{ishi2}, formulae (56), (79), (90), the vector type electromagnetic fields and the vector type gravitational fluctuations obey to this equation with $\lambda=-3$, and the tensor type gravitational fluctuations are solutions with $\lambda= 0$.

The Poincar\'e patch $\mathcal{P}$  is the domain $X_0>X_4$ of the Anti-de Sitter space. We introduce the Poincar\'e coordinates $t,\mathrm{x},z$ defined by :
$$
z=\frac{1}{k^2(X_0-X_4)},\;\;t=kzX_5,\;\;\mathrm{x}^i=kzX_i,\;\;i=1,2,3.
$$
Then the Poincar\'e chart $\mathcal{P}$ of $AdS^5$ is simply
$$
\mathcal{P}:=\RR_t\times\RR_{\mathbf x}^3\times]0,\infty[_z,
$$
endowed with the conformally flat metric
$$
ds^2=\left(\frac{1}{kz}\right)^2\left(dt^2-d{\mathbf x}^2-dz^2\right).
$$
Its time-like horizon $z=0$ is a part of the global boundary $\rho=\frac{\pi}{2}$ (see \cite{bayona} for a complete description). The 4-dimensional Minkowski brane with a negative (resp. positive) tension is the boundary $z=1$ of the bulk $\mathcal{B}_-$ given by $0<z<1$ (resp. $\mathcal{B}_+$ given by $1<z$).
\begin{center}
\begin{tikzpicture}
\useasboundingbox (-1,-2) rectangle (5,9);
\draw[dashed] (4,8) -- (4,0) node[midway,sloped,below] [rotate=180] {$\rho=\frac{\pi}{2}$};
\draw[dashed] (0,0) -- (0,8) node[midway,sloped,above] {$z=0\;\;(\rho=\frac{\pi}{2})$};
\draw (0,0) -- (4,0) node[midway,sloped,below] {$\tau=-\pi$};
\draw (0,8) -- (4,8) node[midway,sloped,above] {$\tau=\pi$};
\draw (0,0) -- (4,4) node[midway,sloped,above] {$t=-\infty,\;z=\infty$};
\draw (4,4) -- (0,8) node[midway,sloped, below] {$t=+\infty,\;z=\infty$};
\draw (0,0) .. controls (2,4) .. (0,8) node[midway,sloped,above] {z=1};
\draw (0.5,4) node {$\mathcal{B}_-$};
\draw (2.5,4) node {$\mathcal{B}_+$};
\fill[fill=black!9!white] (0,0) -- (4,0) -- (4,4) ;
\fill[fill=black!9!white] (4,4) -- (4,8) -- (0,8) ;
\draw (2,-1) node {\it Conformal Penrose diagram of $AdS^5$. $\mathcal{P}$ is the white domain.} ;
\draw (2,-1.5) node {\it
The gray zone is the copy of the Poincar\'e chart defined by $X_0<X_4$.} ;
\end{tikzpicture}
\end{center}

With these nice coordinates, the Klein-Gordon equation is writen as
$$
\left[\square_g+\lambda
k^2\right]u=k^2z^2\left[\partial_t^2-\Delta_{\mathbf{x}}-\partial_z^2+\frac{3}{z}\partial_z+\frac{\lambda}{z^2}\right]u,
$$
that we transform to get the free wave equation on a half 5-dimensional Minkowski space-time, pertubed by a singular potential $\mu z^{-2}$ :
$$
\left[\square_g+\lambda
k^2\right]u=k^2z^{\frac{7}{2}}\left[\partial_t^2-\Delta_{\mathbf{x}}-\partial_z^2+\frac{\mu}{z^2}\right]\left(z^{-\frac{3}{2}}u\right),\;\;\mu=\frac{15}{4}+\lambda.
$$
With $Z_2$ orbifold symmetry imposed across the brane, the boundary condition for $u$ is Neumann on the brane, $\partial_zu=0$, hence for $\Phi=:z^{-\frac{3}{2}}u$, this constraint becomes of Robin type :
$$
\partial_z\Phi(t,\mathrm{x},1)+\frac{3}{2}\Phi(t,\mathrm{x},1)=0.
$$







\begin{thebibliography}{10}

\bibitem{abra}
M. Abramowitz, I.A. Stegun, Eds.
{\it Handbook of Mathematical Functions with formulas, graphs, and mathematical tables}
(Dover, New York, 1965).

\bibitem{ancona}
P. d'Ancona, V. Georgiev, H. Kubo,
Weighted Decay Estimates for the Wave Equation,
{\it J.Differential Equations},
177 (2001), 146-2008.

\bibitem{avis}
S.J. Avis, C.J. Isham, D. Storey,
Quantum field theory in anti-de Sitter space-time,
 {\it Phys. Rev. D},
 18 (10) (1978) : 3565--3576.

\bibitem{DADS}
A. Bachelot,
The Dirac System on the Anti-de Sitter Universe,
{\it Commun. Math. Phys.}
283 (2008), 127--167.

\bibitem{RS}
A. Bachelot,
Wave Propagation and Scattering for the $RS2$ Brane Cosmology Model,
{\it J.H.D.E.}
6(4) (2009), 809--861.

\bibitem{equipart}
A. Bachelot,
Equipartition de l'\'energie pour les syst\`emes
  hyperboliques et formes compatibles,
{\it Ann. Inst. Henri Poincar\'e - Physique th\'eorique}
  46(1) (1987), 45--76.

\bibitem{bayona}
C. A. Ball\'on Bayona, N. R. F. Braga,
Anti-de Sitter boundary in Poincar\'e coordinates,
{\it Gen. Relativity Gravitation}
  39(9) (2007), 1367--1379.

\bibitem{breit1}
P. Breitenlohner, D. Z. Freedman,
Stability in gauged extended supergravity,
{\it Ann. Phys.},
144, 2 (1982), 249--281.

\bibitem{choquet}
Y. Choquet-Bruhat, D .Christodoulou
Elliptic systems in $H_{s,\delta}$ spaces on manifolds which are Euclidean at infinity,
{\it Acta Math.},
146 (1981), 129–150.

\bibitem{deny}
J. Deny, J-L. Lions,
Les espaces du type Beppo Levi,
{\it Ann. Inst. Fourier, Grenoble}
  5 (1955), 305--370.

\bibitem{everitt}
W.N. Everitt, H. Kalf,
The Bessel differential equation and the Hankel transform,
{\it J. Comput. Appl. Math.}
208 (2007), 3--19.

\bibitem{gunther}
P. G\"unther,
{\it Huygens' Principle and Hyperbolic Equations}.
(Academic Press, 1988).

\bibitem{haw}
S. W. Hawking, G. F. R. Ellis,
{\it The large scale structure of space-time}.
(Cambridge University Press, 1973).

\bibitem{hormander1}
L. H\"ormander,
{\it The Analysis of Linear Partial Differential Operators I},
second edition,
(Springer, 1990).


\bibitem{ishi2}
A. Ishibashi, R. M. Wald,
Dynamics in non-globally-hyperbolic, static
    space-times : III. Anti-de-Sitter space-time,
{\it Class. Quantum Grav.}, 21 (2004), 2981--3013.

\bibitem{kalf}
H. Kalf,
A characterization of the Friedrichs extension of Sturm-Liouville operators,
{\it
J. London Math. Soc.}
(2) 17 (1978), 511-521.


\bibitem{keel}
M. Keel, T. Tao,
Endpoint Strichartz Estimates,
{\it Amer. J. Math.}
120 (1998), 955-980.


\bibitem{lions}
J-L.  Lions, E. Magenes,
\newblock {\it Probl\`emes aux limites non homog\`enes et applications I},
(Dunod, 1968).

\bibitem{man}
Ph. D. Mannheim,
{\it Brane-Localized Gravity}
(World Scientific, 2005).

\bibitem{naimark}
M. A. Naimark,
{\it Linear differential operators, Part II},
Frederick Ungar Publishing Co., 1968.

\bibitem{rosenberger}
R. Rosenberger,
\newblock \protect {A new characterization of the Friedrichs extensions of semibounded Sturm-Liouville operators}.
\newblock {\it J. London Math. Soc.}, (2) 31 (1985), 501-510.

\bibitem{olver}
F. W. J.  Olver,
{\it Asymptotics and special functions}, Reprint of the 1974 edition,
(Academic Press,  1997).

\bibitem{titchmarsh}
N. Titchmarsh,
{\it Theory of Fourier integral}, second ed.,
Oxford University Press, 1948.

\bibitem{titchmarsh1}
N. Titchmarsh,
{\it Eigenfunction expansions associated with second-order differential equations},
Oxford University Press, 1946.

\bibitem{torrence}
R. J. Torrence, W. E. Couch,
Transparency of de Sitter and anti-de Sitter spacetimes to multipole fields,
{\it Classical Quantum Gravity}
2 (1985), no. 4, 545--553.

\bibitem{vasy}
A. Vasy,
The wave equation on asymptotically Anti-de Sitter spaces,
arXiv:0911.5440v1,
2009.


\bibitem{vishik}
M. I. Vishik, O. A. Ladyzhenskaya,
Boundary value problems for partial differential equations and certain
classes of operator equations,
{\it  American Mathematical Society Translations}, Ser. 2, Vol. 10 (1958), 223--281.


\bibitem{watson}
G. N. Watson,
{\it A treatise on the theory of Bessel functions},
Reprint of the second (1944) edition (Cambridge University Press,
1995).

\bibitem{wilcox}
C. H. Wilcox,
Initial-Boundary Value Problems for Linear Hyperbolic Partial
  Differential Equations of the Second Order,
{\it  Arch. Rational Mech. Anal.} 10 (1962), 361--400.


\end{thebibliography}
\end{document}